\documentclass[aps,pre,onecolumn,floatfix]{revtex4}

\usepackage{bm}
\usepackage{graphicx}
\usepackage{amssymb,amsfonts,amsmath}
\usepackage{epsf}
\usepackage{color}
\usepackage{hyperref}
\usepackage{subfigure}
\usepackage{epstopdf}
\usepackage{soul}

\usepackage{mathrsfs}

\usepackage{upgreek}

\newcommand{\be}{\begin{equation}}
\newcommand{\ee}{\end{equation}}
\newcommand{\bea}{\begin{eqnarray}}
\newcommand{\eea}{\end{eqnarray}}

\allowdisplaybreaks

\begin{document}

\title{\bf Hydrodynamic correlation and spectral functions of  perfect cubic crystals}

\author{Jo\"el Mabillard}
\email{Joel.Mabillard@ulb.be; \vfill\break ORCID: 0000-0001-6810-3709.}
\author{Pierre Gaspard}
\email{Gaspard.Pierre@ulb.be; \vfill\break ORCID: 0000-0003-3804-2110.}
\affiliation{Center for Nonlinear Phenomena and Complex Systems, Universit{\'e} Libre de Bruxelles (U.L.B.), Code Postal 231, Campus Plaine, B-1050 Brussels, Belgium}

\begin{abstract}
We investigate the collective dynamics of the perfect cubic crystal by deriving from the hydrodynamic equations the time-dependent correlation and the spectral functions characterizing the  fluctuations of mass and momentum densities. We show that the seven hydrodynamic modes of the perfect crystal can be identified from the resonances of these spectral functions. The comparison with those of a fluid is discussed.  Using the numerical values of the thermodynamic, elastic, and transport coefficients computed in our previous paper [J.~Mabillard and P.~Gaspard, arXiv:2311.00757 (2023)]  for a system of hard spheres, the theoretical expressions for the correlation and spectral functions are compared to the same functions directly computed using molecular dynamics simulations. The excellent agreement between theory and simulation provides strong support for the microscopic hydrodynamic theory of perfect crystals based on the local-equilibrium approach. This work sheds light on the fundamental mechanisms governing the collective behavior of matter in the solid state.
\end{abstract}

\maketitle

%%%%%%%%%%%%%%%%%%%%%%%%%%%%%%%%%%%%%%%%%%%%%%%%%%%%%%%%%%%%%%%%%%%
\section{Introduction}
\label{sec:Introduction}
%%%%%%%%%%%%%%%%%%%%%%%%%%%%%%%%%%%%%%%%%%%%%%%%%%%%%%%%%%%%%%%%%%%
The time-dependent  correlation functions and the spectral functions characterizing the  fluctuations of the microscopic densities related to the slow modes of a statistical system play a crucial role in  understanding  its collective dynamics. Phenomena such as the propagation and the attenuation of sound waves or the conduction of heat can be investigated using the resonances of the spectral functions.  In particular, the broadness of the resonances is caused by the damping of the modes due to transport properties such as the viscosities and the heat conductivities, which are sources of irreversibility at the macroscale. The correlation and spectral functions have been extensively studied in fluids~\cite{F75,BP76,BY80}. For instance, the time-dependent correlation function of the mass or particle density is known as the intermediate scattering function and its corresponding spectral function as the dynamic structure factor \cite{BP76,BY80,vH54}.  The latter gives analytical expressions for the cross-sections of light, x-ray, or neutron scattering in fluids and  other phases of condensed matter \cite{vH54,G55,BF66,SBR67,MD69,SROR72,FC76,TES78,AAY83}.

In this paper, we consider perfect cubic crystals.  In addition to the five modes coming from the fundamental conservation laws of energy, momentum, and mass, that are already present in a fluid, a single component crystal has three more modes coming from the breaking of the continuous translational symmetry in the three directions of space.  Perfect crystals contain no vacancy, i.e., all the sites of their lattice are fully occupied.  Consequently, the hydrodynamics of perfect crystals has seven slow modes, because the eighth mode of vacancy diffusion is absent.  The seven hydrodynamic modes are the six longitudinal and transverse sound modes and the diffusive mode of heat conduction.  All these modes can be identified in the correlation and spectral functions of mass and momentum densities, as shown here below.

On the one hand, the correlation and spectral functions can be deduced from the dissipative hydrodynamics of perfect crystals combined with the hypothesis of regression of fluctuations at statistical equilibrium \cite{O31b}. The seven hydrodynamic modes are all clearly identified from the resonances of the spectral functions. The locations and the widths of the peaks of the spectral functions give the speeds and the damping rates of the longitudinal and transverse sound waves, as well as the diffusivity of the heat mode. On the other hand, these functions can be directly computed with molecular dynamics simulations performed over large enough spatiotemporal scales to reach the hydrodynamic regime. Using the values of the thermodynamic, elastic, and transport coefficients that we have previously obtained in reference~\cite{MG23_primo},  the analytical expressions of the correlation and spectral functions can be  compared to their numerical computations that are directly obtained using the simulation.  To make this comparison possible, the computations are performed with elastically colliding hard spheres, since the hard-sphere system forms a face-centered cubic (fcc) crystal at high enough density.  The comparison of the results of the two approaches thus provides a test for the predictions of hydrodynamics and the  framework of the local-equilibrium approach for a perfect crystalline solid~\cite{MG20,MG21}, in the same way as we have done for fluids in reference~\cite{MG23}.

The paper is organized as follows.  In section~\ref{sec:Correl-Spectr}, the time-dependent correlation functions and their corresponding spectral functions are introduced at the microscopic level of description in terms of the particles composing the system and their motion ruled by Hamiltonian classical mechanics.  These functions are considered to characterize the fluctuations of mass (or particle) and momentum densities with respect to the equilibrium probability distribution.  In section~\ref{sec:HydroSpectrum}, the hydrodynamics of perfect cubic crystals is presented.  The linearized hydrodynamic equations ruling the mass, energy, and momentum densities, and the strain tensor are solved using Fourier-Laplace transforms.  If the wave vector is oriented in special directions of the fcc lattice, the set of equations splits into decoupled longitudinal and transverse equations, which can be solved analytically to obtain the intermediate scattering function and the dynamic structure factor for the longitudinal components and further functions for the transverse components.  The spectral functions have resonance peaks determined by the dispersion relations of the seven hydrodynamic modes of the crystal.  The comparison between the crystal and the fluid is discussed.  The speeds and the acoustic attenuation coefficients of the sound waves, as well as the diffusivity of the heat mode, are evaluated as a function of the particle density using the hydrodynamic properties and, in particular, the three viscosities and the heat conductivity computed for the hard-sphere crystal in our previous paper~\cite{MG23_primo}.  The dispersion relations of the seven hydrodynamic modes are explicitly obtained as a function of the wave number for the crystal.  In section~\ref{sec:MDHS}, the correlation and spectral functions are computed for the perfect hard-sphere crystal with molecular dynamics simulations and compared with the predictions of section~\ref{sec:HydroSpectrum} based on hydrodynamics for different values of the particle density and the wave number.  Conclusion and perspectives are given in section~\ref{sec:conclusion}.

{\it Notations.} The Latin indices $a, b, c, d, \ldots = x, y, z$ correspond to spatial coordinates. The indices $i,j = 1,2,\ldots,N$ are the labels of the atoms/particles. Unless explicitly stated, Einstein's convention of summation over repeated indices is adopted. $k_{\rm B}$ denotes Boltzmann's constant and ${\rm i}=\sqrt{-1}$.

%%%%%%%%%%%%%%%%%%%%%%%%%%%%%%%%%%%%%%%%%%%%%%%%%%%%%%%%%%%%%%%%

\section{Microscopic correlation and spectral functions}
\label{sec:Correl-Spectr}
%%%%%%%%%%%%%%%%%%%%%%%%%%%%%%%%%%%%%%%%%%%%%%%%%%%%%%%%%%%%%%%%

\subsection{General properties}

At the microscale, the motion of the $N$ atoms composing the crystal is ruled by Hamiltonian classical mechanics if the temperature is large enough for quantum effects to be negligible.  In this description, the positions and momenta of the atoms determine the microscopic state of the system in the phase space as $\Gamma=({\bf r}_i,{\bf p}_i)_{i=1}^N\in{\mathbb R}^{6N}$. This state evolves in time according to Hamilton's equations ${\rm d}\Gamma/{\rm d}t=\{\Gamma,H\}$, where $\{\cdot,\cdot\}$ denotes the Poisson bracket and $H(\Gamma)$ the Hamiltonian function.  The integration of Hamilton's equations generates the trajectories of motion $\Gamma_t=\Phi^t\Gamma_0$, mapping the initial conditions $\Gamma_0$ onto the state $\Gamma_t$ at time $t$.  The Hamiltonian dynamics preserves the phase-space volumes, ${\rm d}\Gamma_t={\rm d}\Gamma_0$, which is known as the Liouville theorem.  Furthermore,  the property of {\it microreversibility} may be satisfied, according to which the dynamics is invariant under the time-reversal transformation:  $\Theta({\bf r}_i,{\bf p}_i)=({\bf r}_i,-{\bf p}_i)$.  This is the case if the Hamiltonian function is even under time reversal $H(\Theta\Gamma)=H(\Gamma)$, which implies that the phase-space flow $\Phi^t$ is transformed according to $\Theta\circ\Phi^t=\Phi^{-t}\circ\Theta$.

In order to characterize the dynamical fluctuations of some observable quantities $A(\Gamma)$ and $B(\Gamma)$
around equilibrium, we may introduce their time-dependent correlation function as
\begin{align}
\label{eq:correl-dfn}
C_{AB}(t) \equiv \langle \delta A(0) \, \delta B(t)\rangle_{\rm eq}
\qquad\mbox{with}\qquad
\delta A \equiv A - \langle A\rangle_{\rm eq}
\qquad\mbox{and}\qquad
\delta B \equiv B - \langle B\rangle_{\rm eq} \, ,
\end{align}
where $\langle\cdot\rangle_{\rm eq}$ denotes the statistical average with respect to the equilibrium probability distribution ${\cal P}_{\rm eq}(\Gamma)$.  Accordingly, the time-dependent correlation function can be expressed as
\begin{align}
\label{eq:correl-prob}
C_{AB}(t) = \int {\rm d}\Gamma \, {\cal P}_{\rm eq}(\Gamma)\,  \delta A(\Gamma) \, \delta B(\Phi^t\Gamma) \, .
\end{align}
Providing that the dynamics is mixing, correlation functions are expected to converge to zero in the long-time limit,  $\lim_{t\to\infty} C_{AB}(t)=0$, expressing the loss of memory of the initial conditions over increasing time lapses.  We note that the correlation function at time $t=0$ defines the equilibrium property of covariance between $A$ and $B$: $C_{AB}(0)=\langle \delta A \, \delta B\rangle_{\rm eq}$.  Otherwise, for $t\ne 0$, the correlation function characterizes the nonequilibrium  behavior of these observable quantities when the system is in the thermodynamic equilibrium macrostate.

The equilibrium probability distribution has the property of being stationary with respect to the time evolution ${\cal P}_{\rm eq}(\Phi^t\Gamma)={\cal P}_{\rm eq}(\Gamma)$.  As a consequence, we have that $\langle \delta A(0) \, \delta B(t)\rangle_{\rm eq}=\langle \delta A(\tau) \, \delta B(\tau+t)\rangle_{\rm eq}$ upon the time translation $\tau\in{\mathbb R}$.  Taking $\tau=-t$, the equilibrium stationarity implies that the time-dependent correlation function obeys 
$C_{AB}(t)=C_{BA}(-t)$.

Furthermore, if the property of microreversibility holds (and the total momentum of the particles is equal to zero in the frame that is considered), the equilibrium probability distribution is symmetric under time reversal ${\cal P}_{\rm eq}(\Theta\Gamma)={\cal P}_{\rm eq}(\Gamma)$, because it is expressed in terms of the Hamiltonian function.  Consequently, the time-dependent correlation function satisfies $C_{AB}(t)=\epsilon_A\epsilon_B C_{AB}(-t)$, if the  observable quantities $A$ and $B$ have the parities $\epsilon_A=\pm1$ and $\epsilon_B=\pm1$ under time reversal.

In the case where the quantity $B$ is the complex conjugate of $A$, i.e., $B=A^*$, the aforementioned properties and their combination imply that
\begin{align}
\mbox{equilibrium stationarity:} \qquad & C_{AA^*}(t) = C_{AA^*}^*(-t) \, , \label{eq:C:eq-station}\\
\mbox{microreversibility:} \qquad & C_{AA^*}(t) = C_{AA^*}(-t) \, , \label{eq:C:microrev}\\
\mbox{equilibrium stationarity and microreversibility:} \qquad & C_{AA^*}(t) = C_{AA^*}^*(t) \, , \label{eq:C:eq-station+microrev}
\end{align}
since $\epsilon_{A^*}=\epsilon_A$.  The property of microreversibility~\eqref{eq:C:microrev} thus implies that the correlation function is even under time reversal and combined with equilibrium stationarity that the correlation function is real.

The frequency content of the fluctuations can be characterized by the associated spectral function defined by the temporal Fourier transform of the correlation function as 
\begin{align}
S_{AB}(\omega) \equiv \int_{-\infty}^{+\infty} C_{AB}(t) \, {\rm e}^{-{{\rm i}}\omega t} \, {\mathrm d}t \, .
\end{align}
The property of equilibrium stationarity has for consequence that $S_{AB}(\omega)=S_{BA}(-\omega)$ and microreversibility that $S_{AB}(\omega)=\epsilon_A\epsilon_B S_{AB}(-\omega)$.  Their combination gives $S_{AB}(\omega)=\epsilon_A\epsilon_B S_{BA}(\omega)$.  Therefore, if $B=A^*$, we have that
\begin{align}
\mbox{equilibrium stationarity:} \qquad & S_{AA^*}(\omega) = S_{AA^*}^*(\omega) \, , \label{eq:S:eq-station}\\
\mbox{microreversibility:} \qquad & S_{AA^*}(\omega) = S_{AA^*}(-\omega) \, , \label{eq:S:microrev}\\
\mbox{equilibrium stationarity and microreversibility:} \qquad & S_{AA^*}(\omega) = S_{AA^*}^*(-\omega) \, .\label{eq:S:eq-station+microrev}
\end{align}

In the complex plane, where these spectral functions can be extended to become functions of complex frequencies,  they may have poles or other singularities.  Assuming that the poles are located at the complex frequencies $\omega_r={\rm Re}\, \omega_r + {\rm i}\,{\rm Im}\, \omega_r$,  the real part ${\rm Re}\, \omega_r$ gives the characteristic frequency of the corresponding mode and the imaginary part ${\rm Im}\, \omega_r$ corresponds to the damping rate of the mode.  Therefore, the relaxation time of the mode can be evaluated as $\tau_r=1/\vert{\rm Im}\, \omega_r\vert$.  

Considering the observables $A$ and $B=A^*$ in equation~\eqref{eq:correl-dfn} as Fourier modes of wave vector $\bf q$ in the solid, its hydrodynamic properties such as its slow modes are therefore not only identified but also characterized using their correlation and spectral functions.  In this way, the dispersion relations $\omega_r({\bf q})$ of the hydrodynamic modes can be obtained from the poles of the spectral functions.  A mode is thus diffusive if the real part ${\rm Re}\, \omega_r({\bf q})$ is equal to zero.  Otherwise, the mode is propagating and its propagation speed is given by $c_r=\lim_{q\to 0} \vert {\rm Re}\, \omega_r({\bf q})\vert/q$ with $q=\Vert{\bf q}\Vert$. In the following, we compute the spectral functions required to characterize the seven hydrodynamic modes of the one-component perfect crystalline solid. The eighth mode related to the diffusion of vacancies is much slower than the other modes, and is neglected in a first approximation to the macroscopic description of the crystal.  The spectral functions we consider include the dynamic structure factor characterizing the density fluctuations and the spectral functions of momentum density fluctuations. This derivation generalizes the calculation of these functions for the fluid~\cite{MG23,F75,BP76,BY80}.

\subsection{Characterization of particle or mass density fluctuations}

The microscopic particle density is given by $\hat{n}({\bf r},t)\equiv\sum_{i=1}^N\delta[{\bf r}-{\bf r}_i(t)]$ and its Fourier transform by 
\begin{align}
\label{eq:n(q)}
\hat{n}({\bf q},t)=\int_V \hat n({\bf r},t) \, {\rm e}^{{\rm i}{\bf q}\cdot {\bf r}}\, {\rm d}{\bf r}=\sum_{i=1}^N {\rm e}^{{\rm i} {\bf q}\cdot{\bf r}_i(t)} \, .
\end{align}

We note that, if the dynamics is simulated with periodic boundary conditions in a large cubic domain of size $L$ and volume $V=L^3$, the density has the periodicity $\hat{n}({\bf r},t)=\hat{n}({\bf r}+{\bf L},t)$ with ${\bf L}\equiv L\left(m_x{\bf e}_x+m_y{\bf e}_y+m_z{\bf e}_z \right)$ and $(m_x,m_y,m_z)\in {\mathbb Z}^3$. Accordingly, the wave vector ${\bf q}$ of the Fourier modes defined in this domain should have the form ${\bf q} = (2\pi/L)\left(n_x{\bf e}_x+n_y{\bf e}_y+n_z{\bf e}_z \right)$ with $(n_x,n_y,n_z)\in {\mathbb Z}^3$.

In the crystalline phase, the equilibrium mean value of the particle density $n_{\rm eq}({\bf r})\equiv \langle \hat{n}({\bf r})\rangle_{\rm eq}$ is a periodic function in the three-dimensional space with the periodicity of the crystal lattice.
Therefore, the equilibrium density can be decomposed as $n_{\rm eq}({\bf r})  = \sum_{\bf G} n_{\rm eq,{\bf G}}\,{\rm e}^{-{\rm i}{\bf G}\cdot {\bf r}}$ as a sum over the reciprocal lattice vectors $\bf G$ and the Fourier transform of the equilibrium density is equal to zero unless the wave vector coincides with a reciprocal lattice vector:
\begin{align}
n_{\rm eq}({\bf q}) =\int_V n_{\rm eq}({\bf r}) \, {\rm e}^{{\rm i}{\bf q}\cdot {\bf r}}\, {\rm d}{\bf r}&= 
\left\{
\begin{array}{ll} V\, n_{\rm eq,{\bf G}} \, , & \text{if} \ {\bf q}={\bf G} \, ,\\
0 \, , & \text{otherwise.}   \end{array}
\right.
\label{eq:neqg}
\end{align}

Now, the fluctuations of density are characterized by the time-dependent autocorrelation function of the Fourier components $\hat{n}({\bf q},t)$ of the microscopic particle density, which is called the {\it intermediate scattering function}:
\begin{align}
\label{eq:ISF-dfn}
F({\bf q},t)&\equiv  \frac{1}{N} \langle \delta \hat{n}({\bf q},t)\, \delta \hat{n}^*({\bf q},0)\rangle_{\rm eq}= \frac{1}{N}\langle  \hat{n}({\bf q},t)\, \hat{n}^*({\bf q},0)\rangle_{\rm eq}-\frac{1}{N \, }|n_{\rm eq}({\bf q})|^2 \, ,
\end{align}
where the second equality results from $\delta \hat{n}({\bf q},t)=\hat{n}({\bf q},t)-n_{\rm eq}({\bf q})$.
If we are interested in the hydrodynamic regime, the magnitude of the wave vector $q=\Vert{\bf q}\Vert$ should take its smallest possible nonvanishing values, which are of the order of  $q_{\rm min}\sim2\pi/L$.  In contrast, the smallest nonvanishing values of the magnitude $G=\Vert{\bf G}\Vert$ of the reciprocal lattice vector $\bf G$ are of the order of $G_{\rm min}\sim 2\pi/a$, where $a$ is the size of a lattice cell, which is much smaller than the size $L$ of the simulated system.  As a consequence, we have that $\Vert{\bf q}\Vert \ll \Vert{\bf G}\Vert$ and $n_{\rm eq}({\bf q}) =0$ because of equation~\eqref{eq:neqg}.  Under such circumstances, the second term in the right-hand side of equation~\eqref{eq:ISF-dfn} does not contribute to the intermediate scattering function, which can thus be obtained from
\begin{align}
\label{eq:ISFMD}
F({\bf q},t)&= \frac{1}{N} \langle \hat{n}({\bf q},t)\, \hat{n}^*({\bf q},0)\rangle_{\rm eq} = \frac{1}{N} \left\langle\sum_{i,j=1}^N{\rm e}^{{\rm i} {\bf q}\cdot\left[{\bf r}_i(t)-{\bf r}_j(0)\right]}\right\rangle_{\rm eq}
\end{align}
for ${\bf q}\ne 0$. 

The so-called {\it dynamic structure factor} is defined as the associated spectral function according to
\begin{align}
\label{eq:DSF-dfn}
S({\bf q},\omega) \equiv \int_{-\infty}^{+\infty} F({\bf q},t) \, {\rm e}^{-{{\rm i}}\omega t} \, {\mathrm d}t \, .
\end{align}
The {\it static structure factor} is defined by the intermediate scattering function at time $t=0$: $S({\bf q})\equiv F({\bf q},0)=\int_{-\infty}^{+\infty} S({\bf q},\omega) \, {\rm d}\omega/(2\pi)$.
 
 We note that these functions can be equivalently expressed in terms of the mass density, which is defined as $\hat\rho\equiv m \hat n$ with the mass~$m$ of the particles.  In particular, the intermediate scattering function~\eqref{eq:ISF-dfn} also reads 
 \begin{align}
\label{eq:ISF-mass}
F({\bf q},t)&\equiv  \frac{1}{Nm^2} \langle \delta \hat{\rho}({\bf q},t)\, \delta \hat{\rho}^*({\bf q},0)\rangle_{\rm eq} \, .
\end{align}

\subsection{Characterization of momentum density fluctuations}

The microscopic momentum density is defined by $\hat g^a({\bf r},t)\equiv\sum_{i=1}^N p_i^a(t)\, \delta[{\bf r}-{\bf r}_i(t)]$, so that its Fourier modes are given by
\begin{align}
\label{eq:g(q)}
\hat g^a({\bf q},t) = \int_V \hat g^a({\bf r},t) \, {\rm e}^{{\mathrm i}{\bf q}\cdot{\bf r}} \, {\mathrm d}{\bf r} = \sum_{i=1}^N p_i^a(t) \, {\rm e}^{{\mathrm i}{\bf q}\cdot{\bf r}_i(t)} \, .
\end{align}

We introduce the orthonormal basis $\{{\bf e}_{\rm l},{\bf e}_{{\rm t}_1},{\bf e}_{{\rm t}_2}\}$, where ${\bf e}_{\rm l}\equiv{\bf q}/q$ is the unit vector oriented in the direction of the wave vector ${\bf q}$, while the unit vectors ${\bf e}_{{\rm t}_1}$ and ${\bf e}_{{\rm t}_2}$ are oriented in two orthogonal directions perpendicular to the wave vector ${\bf q}$, whereupon these vectors satisfy ${\bf e}_{\sigma}\cdot{\bf e}_{\sigma^\prime}=\delta_{\sigma\sigma^\prime}$, where $\sigma,\sigma^\prime\in\{{\rm l},{\rm t}_1,{\rm t}_2\}$. The longitudinal and transverse components of the momentum ${\bf p}_i(t)$ of a particle are thus given by $p_{\sigma i}(t)={\bf e}_{\sigma}\cdot{\bf p}_i(t)=e_{\sigma}^a \, p_i^a(t)$.  We may introduce similar components for the Fourier modes~\eqref{eq:g(q)} of the microscopic momentum density as $\hat g_{\sigma}({\bf q},t)=e_{\sigma}^a\,\hat{g}^a({\bf q},t)$ for $\sigma={\rm l},{\rm t}_1,{\rm t}_2$.

Using that $\delta \hat{g}^a=\hat{g}^a$ since $\langle\hat g^a\rangle_{\rm eq}=0$, the momentum  density correlation functions are defined by
\begin{align}
\label{eq:mom_cf}
C_{\sigma}({\bf q},t) \equiv \frac{1}{N m^2}\langle\hat g_{\sigma}({\bf q},t)\, \hat g_{\sigma}^*({\bf q},0)\rangle_{\rm eq} =\frac{1}{Nm^2} \left\langle \sum_{i,j=1}^N p_{\sigma i}(t)\, p_{\sigma j}(0) \, {\rm e}^{{\mathrm i}{\bf q}\cdot\left[{\bf r}_i(t)-{\bf r}_j(0)\right]} \right\rangle_{\rm eq} 
\end{align}
with $\sigma\in\{{\rm l},{\rm t}_1,{\rm t}_2\}$, and there is no Einstein’s summation for the  indices $\sigma$. The corresponding spectral functions are defined by
\begin{align}
\label{eq:spctr_fns-dfn}
J_{\sigma}({\bf q},\omega) \equiv \int_{-\infty}^{+\infty} C_{\sigma}({\bf q},t) \, {\rm e}^{-{{\rm i}}\omega t} \, {\mathrm d}t \, .
\end{align}

The longitudinal momentum density correlation function is related to the intermediate scattering function~(\ref{eq:ISF-dfn}) according to
\begin{align}
\label{eq:lca_hydro}
C_{\rm l}({\bf q},t) = -\frac{1}{q^2}\frac{{\mathrm d}^2}{{\mathrm d}t^2}F({\bf q},t) \, ,
\end{align}
so that the associated longitudinal spectral function can be expressed as
\begin{align}
\label{eq:lsa_hydro}
J_{\rm l}({\bf q},\omega)=\frac{\omega^2}{q^2}S({\bf q},\omega)
\end{align}
in terms of the dynamic structure factor~\eqref{eq:DSF-dfn}.

Since we consider the fluctuations of the microscopic momentum density around equilibrium, we may introduce a corresponding fluctuating velocity field as $\hat v_\sigma({\bf q},t)\equiv \hat g_\sigma({\bf q},t)/\rho$, where $\rho=\frac{m}{v}\int_v n_{\rm eq}({\bf r})\, {\rm d}{\bf r}$ denotes the spatially averaged equilibrium mass density, which is obtained by integrating the equilibrium particle density $n_{\rm eq}({\bf r})$ over the volume $v$ of the primitive unit cell of the lattice.  Accordingly, the correlation functions~\eqref{eq:mom_cf} read
\begin{align}
\label{eq:v_cf}
C_{\sigma}({\bf q},t) = \frac{N}{V^2}\langle \delta\hat{v}_{\sigma}({\bf q},t)\, \delta\hat{v}_{\sigma}^*({\bf q},0)\rangle_{\rm eq} \, ,
\end{align}
given that $\delta\hat{v}_\sigma=\hat{v}_\sigma$ since $\langle\hat{v}_\sigma\rangle_{\rm eq}=0$, and there is no Einstein’s summation for the  indices $\sigma$.

%%%%%%%%%%%%%%%%%%%%%%%%%%%%%%%%%%%%%%%%%%%%%%%%%%%%%%%%%%%%%%%%
\section{Hydrodynamics of perfect cubic crystals}
\label{sec:HydroSpectrum}
%%%%%%%%%%%%%%%%%%%%%%%%%%%%%%%%%%%%%%%%%%%%%%%%%%%%%%%%%%%%%%%%

%%%%%%%%%%%%%%%%%%%%%%%%%%%%%%%%%%%%%%%%%%%%%%%%%%%%%%%%%%%%%%%%
\subsection{The linearized hydrodynamic equations}
\label{sec:linhydroequ}
%%%%%%%%%%%%%%%%%%%%%%%%%%%%%%%%%%%%%%%%%%%%%%%%%%%%%%%%%%%%%%%%

Perfect crystals are defined as crystals without vacancies in the occupancy of their lattice sites.  Therefore, the eighth mode of vacancy diffusion is absent in such crystals, which have only seven hydrodynamic modes. On large spatiotemporal scales, the hydrodynamics of crystals rules the time evolution of the macroscopic fields that are the mean mass density $\rho({\bf r},t)$, the mean internal energy density $\epsilon_0({\bf r},t)$, the velocity $v^a({\bf r},t)$, and the strain tensor $u^{ab}\equiv(\nabla^a u^b+\nabla^b u^a)/2=u^{ba}$, where $u^a({\bf r},t)$ is the displacement field.  After relaxation, the crystal reaches the equilibrium macrostate, where the macrofields take constant and uniform values.  For the mass density, this value is equal to $\rho=\frac{m}{v}\int_v n_{\rm eq}({\bf r})\, {\rm d}{\bf r}$, as aforementioned.  For the internal energy density and the hydrostatic pressure, these values are respectively denoted $\epsilon_0$ and $p$.  For the velocity and the strain tensor, they are equal to zero.
Around this equilibrium rest macrostate of the crystal, the time evolution of the local deviations $(\delta\rho,\delta\epsilon_0,\delta v^b,\delta u^{ab})$ of the macrofields with respect to their equilibrium values is ruled by the linearized hydrodynamic equations.

Since the perfect hard-sphere crystal is cubic, the linearized set of hydrodynamic equations reads~\cite{MG23_primo,MG21}
 \begin{align}
\partial_t  \delta \rho & = - \rho \, \nabla^a  \delta v^a  \, , \label{macro-eq-rho}\\
 \partial_t \delta \epsilon_0  & = -(\epsilon_0+p) \, \nabla^a \delta v^a + \kappa \, \nabla^2 \delta T \, , \label{macro-eq-e0}\\
\rho\, \partial_t \delta v^b  & = \nabla^a  \delta\sigma^{ab} + \eta^{abcd}\, \nabla^a\nabla^c \delta v^d  \, , \label{macro-eq-v}\\
\partial_t  \delta u^{ab} & = \frac{1}{2}\left(\nabla^a \delta v^b +\nabla^b\delta v^a\right) , \label{macro-eq-u}
\end{align}
where the elastic properties of the crystal are given by the dependence of the reversible stress tensor $\sigma^{ab}$ on the strain tensor $u^{ab}$ and the temperature $T$, and the transport properties by the heat conductivity $\kappa$ and the viscosities $\eta^{abcd}$.  For the perfect hard-sphere crystal, all these equilibrium and nonequilibrium properties have been computed from the microscopic dynamics in reference~\cite{MG23_primo}.

In perfect crystals, where the vacancy concentration can be neglected, we note that the deviation of the trace of the strain tensor $\delta u^{aa}=\nabla^a \delta u^a$ is related to the deviation of the mass density $\delta\rho$ according to $\delta u^{aa}=- \rho^{-1} \delta\rho$, which is consistent with equations~\eqref{macro-eq-rho} and~\eqref{macro-eq-u}~\cite{MG23_primo}.

We introduce the specific internal energy $e\equiv\epsilon_0/\rho$ such that $\delta \epsilon_0 = e\delta \rho + \rho \delta e$ and satisfying the Gibbs relation $\delta e = T\delta s + p\delta \rho/\rho^2$ in terms of the specific entropy $s$, i.e., the entropy per unit mass.  Using equations~\eqref{macro-eq-rho} and~\eqref{macro-eq-e0}, we obtain the evolution equation for the specific entropy as
\begin{align}
\rho\, T\, \partial_t s & =  \kappa\, \nabla^2 \delta T \, .\label{macro-eq-s} 
\end{align}

In order to consider statistically independent fluctuating fields, we perform the change of variables from $(\delta s, \delta v^a, \delta u^{ab})$ to $(\delta T, \delta v^a, \delta u^{ab})$ and we close the system of equations using
\begin{align}
\delta s &=\left(\frac{\partial s}{\partial u^{ab}}\right)_{T,\{u^{cd}\}_{cd \ne ab}} \delta u^{ab} + \left(\frac{\partial s}{\partial T}\right)_{\{u^{ab}\}} \delta T \, , \label{eq:ds} \\
\delta \sigma^{ab} &=\left(\frac{\partial \sigma^{ab}}{\partial u^{cd}}\right)_{T,\{u^{ef}\}_{ef \ne cd}} \delta u^{cd} + \left(\frac{\partial \sigma^{ab}}{\partial T}\right)_{\{u^{cd}\}} \delta T \, . \label{eq:dsigma}
\end{align}
For cubic crystals, the coefficients of these equations can be expressed in terms of the equilibrium thermodynamic properties~\cite{W98} that are the specific heat capacities at constant volume $c_{v}\equiv T(\partial s/\partial T)_{v}$ and constant pressure $c_{p}\equiv T(\partial s/\partial T)_{p}$, their ratio $\gamma\equiv c_p/c_v$, 
\begin{align}
\mbox{the thermal expansion coefficient} \qquad & \alpha \equiv - \frac{1}{\rho} \left(\frac{\partial \rho}{\partial T}\right)_p , \\
\mbox{the isothermal bulk modulus} \qquad & B_T\equiv\rho\left(\frac{\partial p}{\partial \rho}\right)_T ,
\end{align}
and the rank-four tensor of 
\begin{align}
&\mbox{the isothermal stress-strain coefficients} \qquad B^{abcd}_T\equiv\left(\frac{\partial \sigma^{ab}}{\partial u^{cd}}\right)_{T,\{u^{ef}\}_{ef \ne cd}} .
\end{align}

The rank-two tensors satisfy the following Maxwell relations \cite{W98},
\begin{align}
\left(\frac{\partial s}{\partial u^{ab}}\right)_{T,\{u^{cd}\}_{cd \ne ab}} &= - \frac{1}{\rho} \left(\frac{\partial \sigma^{ab}}{\partial T}\right)_{\{u^{cd}\}} .
\end{align}
Moreover, in cubic crystals, they are diagonal and we have
\begin{align}
\left(\frac{\partial \sigma^{ab}}{\partial T}\right)_{\{u^{cd}\}}  &= -\left(\frac{\partial p}{\partial T}\right)_{\rho}  \delta^{ab} =- \alpha B_T\, \delta^{ab} \, ,
\end{align}
so that
\begin{align}
\left(\frac{\partial s}{\partial u^{ab}}\right)_{T,\{u^{cd}\}_{cd \ne ab}} &=\frac{c_v(\gamma-1)}{T \alpha} \, \delta^{ab} \, ,
\end{align}
because of the well-known thermodynamic relation $c_p-c_v=T\alpha^2 B_T/\rho$ \cite{W98}.
Consequently, equations~\eqref{eq:ds} and~\eqref{eq:dsigma} become
\begin{align}
\delta s &=\frac{c_v}{T} \left(\frac{\gamma -1}{\alpha}\, \delta u^{aa}+\delta T\right) , \label{eq:ds2}
\\
\delta \sigma^{ab} & =B^{abcd}_T \, \delta u^{cd}- \alpha B_T \, \delta^{ab} \, \delta T\, . \label{eq:dsigma2}
\end{align}

Since the strain and stress tensors are symmetric, i.e., $u^{ab}=u^{ba}$ and $\sigma^{ab}=\sigma^{ba}$, the rank-four tensors have the symmetries $B^{abcd}_T=B^{badc}_T$ and $\eta^{abcd}=\eta^{badc}$.  If the externally applied stress is isotropic, the isothermal stress-strain tensor has the additional symmetry $B^{abcd}_T=B^{cdab}_T$ \cite{W98}.  Moreover, the viscosity tensor obeys Onsager's reciprocal relations $\eta^{abcd}=\eta^{cdab}$ because of microreversibility \cite{MG21}.  We also note that, for cubic crystals, the rank-four tensors can be expressed in terms of three coefficients, which read  $B_{11}^T$, $B_{12}^T$, and $B_{44}^T$ for the isothermal stress-strain tensor, and $\eta_{11}$, $\eta_{12}$, and $\eta_{44}$ for the viscosity tensor in Voigt's notations.  Also for cubic crystals, the isothermal bulk modulus can be expressed in terms of the isothermal stress-strain coefficients as $B_T=(B_{11}^T+2B_{12}^T)/3$ \cite{W98}.

Now, the substitution of equations~\eqref{eq:ds2} and~\eqref{eq:dsigma2} into equations~\eqref{macro-eq-v},~\eqref{macro-eq-u}, and~\eqref{macro-eq-s} before taking their Fourier-Laplace transform (as defined in appendix~\ref{app:FLt}) gives the following set of equations,
\begin{align}
\left( z +\frac{\kappa }{\rho c_v} q^2\right)\delta \tilde T(\mathbf{q},z)+ z\frac{\gamma -1}{\alpha}\, \delta \tilde u^{aa}(\mathbf{q},z) & =\delta T(\mathbf{q},0)+\frac{\gamma -1}{\alpha}\, \delta u^{aa}(\mathbf{q},0)   \, ,\label{eq:fl-T}\\
 \left( z \rho \delta ^{bd}+\eta^{abcd} q^a q^c\right)  \delta \tilde v^d(\mathbf{q},z) +{\rm i} B^{abcd}_T q^a \delta \tilde u^{cd}(\mathbf{q},z)- {\rm i}  \alpha B_T q^b \delta \tilde T(\mathbf{q},z)  &= \rho\,  \delta v^b(\mathbf{q},0) \, ,\label{eq:fl-v}\\
z \,\delta \tilde u^{ab}(\mathbf{q},z)+  \frac{{\rm i}}{2}\left[q^a \delta \tilde v^b(\mathbf{q},z)+q^b\delta \tilde v^a(\mathbf{q},z)\right] &= \delta u^{ab}(\mathbf{q},0)\, .\label{eq:fl-u} 
\end{align}

The equations \eqref{eq:fl-T}-\eqref{eq:fl-u} can be split into two independent sets composed of three longitudinal and four transverse equations in special directions to be determined for the wave vector $\bf q$. To this end, we first consider the orthonormal basis $\{{\bf e}_{\sigma}\}$ with $\sigma\in\{{\rm l},{\rm t}_1,{\rm t}_2\}$, which has been introduced here above and such that ${\bf q}=q\, {\bf e}_{\rm l}$, where $q=\Vert{\bf q}\Vert$. In this basis, the Fourier transform of the velocity field can be expressed as
\begin{align}
\label{eq:vlt}
{\bf v}({\bf q}) & = \sum_\sigma v_{\sigma}({\bf q})\,{\bf e}_{\sigma}
\qquad\mbox{with}\qquad
v_{\sigma}({\bf q}) = {\bf e}_\sigma \cdot {\bf v}({\bf q}) \, .
\end{align}
Similarly, the Fourier transform of the displacement field reads
\begin{align}
\label{eq:ult}
{\bf u}({\bf q}) & = \sum_\sigma u_{\sigma}({\bf q})\,{\bf e}_{\sigma}
\qquad\mbox{with}\qquad
u_{\sigma}({\bf q}) = {\bf e}_\sigma \cdot {\bf u}({\bf q}) \, .
\end{align}
Since the spatial Fourier transform has the effect of replacing the gradient $\boldsymbol{\nabla}$ by $-{\rm i}{\bf q}$ and because the wave vector can be expressed as ${\bf q}=q\, {\bf e}_{\rm l}$, the strain tensor $u^{ab}=(\nabla^a u^b+\nabla^b u^a)/2$ is transformed into
\begin{align}
\label{eq:uablt}
u^{ab}(\mathbf{q})&= -\frac{\rm i}{2} \left[q^a u^b({\bf q}) + q^b u^a({\bf q}) \right] = -\frac{\rm i}{2}\, q \sum_\sigma u_\sigma({\bf q}) \left(e^a_{\rm l} \, e^b_{\sigma} + e^b_{\rm l} \, e^a_{\sigma}\right) ,
\end{align}
where
\begin{align}
u_{\rm l}({\bf q}) & = {\rm i}\, q^{-1} \,  u^{ab}({\bf q}) \, e^a_{\rm l} \, e^b_{\rm l}
\qquad\mbox{and}\qquad
u_{{\rm t}_k}({\bf q}) = 2\, {\rm i}\, q^{-1} \, u^{ab}({\bf q}) \, e^a_{\rm l} \, e^b_{{\rm t}_k}
\end{align}
for $k=1,2$ \cite{FC76}.  Moreover, we have that $u^{aa}({\bf q}) = -{\rm i}\, q\, u_{\rm l}({\bf q})$, since ${\bf e}_{\rm l}\cdot{\bf e}_{\sigma}=\delta_{{\rm l}\sigma}$.

Contracting equation~\eqref{eq:fl-v} with $e^b_{\sigma}$, the rank-four tensors of isothermal stress-strain coefficients and viscosities lead to the following rank-two tensors,
\begin{align}
\label{eq:3x3_B+eta}
B_{\sigma\sigma^\prime}^T & \equiv \left(B^{abcd}_T e^a_{\rm l} e^c_{\rm l}\right) e^b_{\sigma} \, e^d_{\sigma^\prime}
\qquad\mbox{and}\qquad
\eta_{\sigma\sigma^\prime} \equiv \left(\eta^{abcd} e^a_{\rm l} e^c_{\rm l}\right) e^b_{\sigma} \, e^d_{\sigma^\prime} \, .
\end{align}
As shown in appendix~\ref{app:calc_tle}, the remarkable property is that these rank-two tensors can be simultaneously diagonalized if the wave vector $\bf q$ is oriented in the special directions of table~\ref{Tab:LTCoeffs} with respect to the axes of the cubic crystal (and symmetry-related directions).  In these special directions, we thus have that $B_{\sigma\sigma^\prime}^T=B_{\sigma}^T\delta_{\sigma\sigma^\prime}$ and $\eta_{\sigma\sigma^\prime}=\eta_{\sigma}\delta_{\sigma\sigma^\prime}$ with the coefficients $B_{\sigma}^T$ and $\eta_{\sigma}$ reported in table~\ref{Tab:LTCoeffs} for $\sigma\in\{{\rm l},{\rm t}_1,{\rm t}_2\}$.

%%%%%%%%%%%%%%%%%%%%%%%%%%%%%%%%%%%%%%%%%%%%%%%%%%%%%%%%%%

\begin{table}[h!]
\begin{tabular}{c @{\hskip 1cm} c @{\hskip 1cm} c @{\hskip 1cm} c  }
\hline\hline
Direction&   $[100]$     &    $[110]$  &  $[111]$   	 \\
\hline  
${\bf e}_{\rm l}$ & ${\bf e}_x$  	& $({{\bf e}_x+{\bf e}_y})/{\sqrt{2}}$ 	& $({{\bf e}_x+{\bf e}_y+{\bf e}_z})/{\sqrt{3}}$	\\
$ {\bf e}_{{\rm t}_1}$		& $  {\bf e}_y$		& $({{\bf e}_x-{\bf e}_y})/{\sqrt{2}}$ & $ ({{\bf e}_x-{\bf e}_y})/{\sqrt{2}}$ \\
$ {\bf e}_{{\rm t}_2}$		& $  {\bf e}_z$		& $  {\bf e}_z$ & $({{\bf e}_x+{\bf e}_y-2{\bf e}_z})/{\sqrt{6}}$ \\
$B^T_{\rm l} $		& $B^T_{11}$		& $({B^T_{11}+B^T_{12}+2B^T_{44}})/{2}$ & $ ({B^T_{11}+2B^T_{12}+4B^T_{44}})/{3}$ \\
$ B^T_{{\rm t}_1}$		& $B^T_{44}$		& $({B^T_{11}-B^T_{12}})/{2}$ & $ ({B^T_{11}-B^T_{12}+B^T_{44}})/{3}$ \\
$ B^T_{{\rm t}_2}$		& $B^T_{44}$		& $B^T_{44}$ & $ ({B^T_{11}-B^T_{12}+B^T_{44}})/{3}$ \\
$\eta_{\rm l} $		& $ \eta_{11}$		& $({\eta_{11}+\eta_{12}+2\eta_{44}})/{2}$ & $({\eta_{11}+2\eta_{12}+4\eta_{44}})/{3}$ \\
$ \eta_{{\rm t}_1}$		& $\eta_{44}$		& $({\eta_{11}-\eta_{12}})/{2}$ & $({\eta_{11}-\eta_{12}+\eta_{44}})/{3}$ \\
$ \eta_{{\rm t}_2}$		& $\eta_{44}$		& $\eta_{44}$ & $({\eta_{11}-\eta_{12}+\eta_{44}})/{3}$ \\                
\hline\hline
\end{tabular}
\caption{The stress-strain coefficients $B^T_{\sigma}$ and the viscosity coefficients $\eta_{\sigma}$ in the longitudinal and transverse directions ${\bf e}_\sigma$ with $\sigma\in\{{\rm l},{\rm t}_1,{\rm t}_2\}$ for the wave vector ${\bf q}$ oriented in the directions $[100]$, $[110]$, and $[111]$ of the cubic crystal, as expressed using Voigt's notations.  See appendix~\ref{app:calc_tle} for their calculations.}
\label{Tab:LTCoeffs}
\end{table}

%%%%%%%%%%%%%%%%%%%%%%%%%%%%%%%%%%%%%%%%%%%%%%%%%%%%%%%%%%

Therefore, contracting equation~\eqref{eq:fl-v} with $e^b_{\rm l}$ and equation~\eqref{eq:fl-u} with $e^a_{\rm l}e^b_{\rm l}$, and using $\delta\rho=-\rho\, \delta u^{aa}$, we obtain from equations~\eqref{eq:fl-T}-\eqref{eq:fl-u} the following set of longitudinal equations,
\begin{align}
z\, \delta \tilde \rho(\mathbf{q},z)- {\rm i} \rho  q \, \delta \tilde v_{\rm l}(\mathbf{q},z)&=\delta \rho(\mathbf{q},0)\, ,\label{eq:zrho_l}\\
\left( z +\gamma D_T q^2\right) \delta \tilde T(\mathbf{q},z)- {\rm i} \frac{\gamma-1}{\alpha} q\, \delta \tilde v_{\rm l}(\mathbf{q},z) & =\delta T(\mathbf{q},0)   \,, \label{eq:zT_l}\\
\left( z +D_vq^2\right)  \delta \tilde v_{\rm l}(\mathbf{q},z) - {\rm i} \frac{B^T_{\rm l}}{\rho^2}  q \, \delta \tilde \rho (\mathbf{q},z)- {\rm i} \frac{\alpha B_T}{\rho} q \, \delta \tilde T(\mathbf{q},z)  &=   \delta v_{\rm l}(\mathbf{q},0) \,, \label{eq:zv_l}
\end{align}
where the  longitudinal kinematic viscosity is defined as  $D_v\equiv \eta_{\rm l}/\rho$ and the thermal diffusivity  as $D_T\equiv \kappa/(\rho c_p)$. 

Furthermore, contracting equation~\eqref{eq:fl-v} with $e^b_{{\rm t}_k}$ and equation~\eqref{eq:fl-u} with $e^a_{\rm l}e^b_{{\rm t}_k}$, we find the decoupled sets of the two following transverse equations,
\begin{align}
\left( z +\frac{\eta_{{\rm t}_k}}{\rho}q^2\right)  \delta \tilde v_{{\rm t}_k}(\mathbf{q},z) +  \frac{B^T_{{\rm t}_k}}{\rho} q^2 \delta \tilde u_{{\rm t}_k} (\mathbf{q},z) &=  \delta v_{{\rm t}_k}(\mathbf{q},0) \, , \label{eq:zv_t}\\		
z\, \delta \tilde u_{{\rm t}_k}(\mathbf{q},z)- \delta \tilde v_{{\rm t}_k}(\mathbf{q},z)&=\delta u_{{\rm t}_k}(\mathbf{q},0)\label{eq:zu_t}\, ,	
\end{align}
where $k=1,2$. The coefficients $B^T_{\rm l}$,  $B^T_{{\rm t}_k}$,  $\eta_{\rm l}$, and  $\eta_{{\rm t}_k}$ depend on the direction of ${\bf q}$ and are given in table~\ref{Tab:LTCoeffs}.

The details of these calculations are given in appendix~\ref{app:calc_tle}.

%%%%%%%%%%%%%%%%%%%%%%%%%%%%%%%%%%%%%%%%%%%%%%%%%%%%%%%%%%%%%%%%
\subsection{Longitudinal correlation and spectral functions} 
\label{sec:longCSF}
%%%%%%%%%%%%%%%%%%%%%%%%%%%%%%%%%%%%%%%%%%%%%%%%%%%%%%%%%%%%%%%%

The set of longitudinal equations can be cast in a matrix form as
\begin{align}
\label{eq:M-phi=phi}
\boldsymbol{\mathsf M}({q},z) \cdot \delta\tilde {\boldsymbol{\phi}}({\bf q},z)= \delta\boldsymbol{\phi}({\bf q},0) \, ,
\end{align}
where $\delta\boldsymbol{\phi}=(\delta\rho,\delta T,\delta v_{\rm l})^{\rm T}$ and 
\begin{align}
\boldsymbol{\mathsf M}({q},z) \equiv \left[
\begin{array}{lll}
z &  0  & -{\rm i}\rho q  \\
0 & z+\gamma D_T q^2 & -{\rm i} \frac{\gamma-1}{\alpha } q\\
- {\rm i}  \frac{B^T_{\rm l}}{\rho^2}q &  -{\rm i} \frac{\alpha B_T}{\rho} q & z+D_vq^2 \\
\end{array}
\right] .
\end{align}
On the basis of the hypothesis of regression of fluctuations~\cite{MG23},  the deviations $\delta\boldsymbol{\phi}$ of the fields  can be replaced by their microscopic expressions $\delta\boldsymbol{\hat\phi}$. As a consequence, the Laplace transforms of the correlation functions can be obtained by solving the so-modified equation~(\ref{eq:M-phi=phi}) to get the fluctuating fields $\delta\hat{\tilde{\boldsymbol{\phi}}}({\bf q},z)$, multiplying them by ${\delta\boldsymbol{\hat\phi}}^{*{\rm T}}({\bf q},0)$, and taking the statistical average $\langle\cdot\rangle_{\rm eq}$ with respect to the equilibrium probability distribution.  There is no coupling between the matrix elements of $\boldsymbol{\mathsf M}^{-1}$, since the equal-time correlation matrix $\langle{\delta\boldsymbol{\hat\phi}}({\bf q},0)\,{\delta\boldsymbol{\hat\phi}}^{*{\rm T}}({\bf q},0)\rangle_{\rm eq}$ is diagonal because the fluctuating fields ${\delta\boldsymbol{\hat\phi}}$ are statistically independent. The Laplace transform of the correlation functions for density, temperature, and longitudinal momentum density are thus given by
\begin{align}
\frac{\tilde F({\bf{q}},z)}{F({\bf q},0)}= \frac{\langle   \delta\hat{\tilde \rho}({\bf{q}},z)  \,  \delta\hat{\rho}^*({\bf{q}},0)  \rangle_{\rm eq}}{\langle   \delta\hat{\rho}({\bf{q}},0)  \,  \delta\hat{\rho}^*({\bf{q}},0)  \rangle_{\rm eq}}&=\frac{ ( z+D_vq^2)(z+\gamma D_T q^2)+\left(\gamma-1\right) B_Tq^2/\rho}{\det \boldsymbol{\mathsf M}({q},z)}\, ,\label{eq:Sks}\\
\frac{\langle   \delta \hat{\tilde T}({\bf{q}},z)    \delta \hat{T}^*({\bf{q}},0)  \rangle_{\rm eq}}{\langle   \delta \hat{T}({\bf{q}},0)    \delta \hat{T}^*({\bf{q}},0)  \rangle_{\rm eq}}&=\frac{z( z+D_vq^2)+B^T_{\rm l}q^2/\rho}{\det \boldsymbol{\mathsf M}({q},z)}\, ,\\
\frac{\langle   \hat{\tilde v}_{\rm l}({\bf{q}},z)    \hat{v}_{\rm l}^*({\bf{q}},0)  \rangle_{\rm eq}}{\langle   \hat{v}_{\rm l}({\bf{q}},0)  \hat{v}^*_{\rm l}({\bf{q}},0)  \rangle_{\rm eq}}&=\frac{ z(z+\gamma D_T q^2)}{\det \boldsymbol{\mathsf M}({q},z)}\, .
\end{align}

\begin{figure}[h!] \centering
    \begin{minipage}{0.5\textwidth}
        \centering
        \includegraphics[width=1.\textwidth]{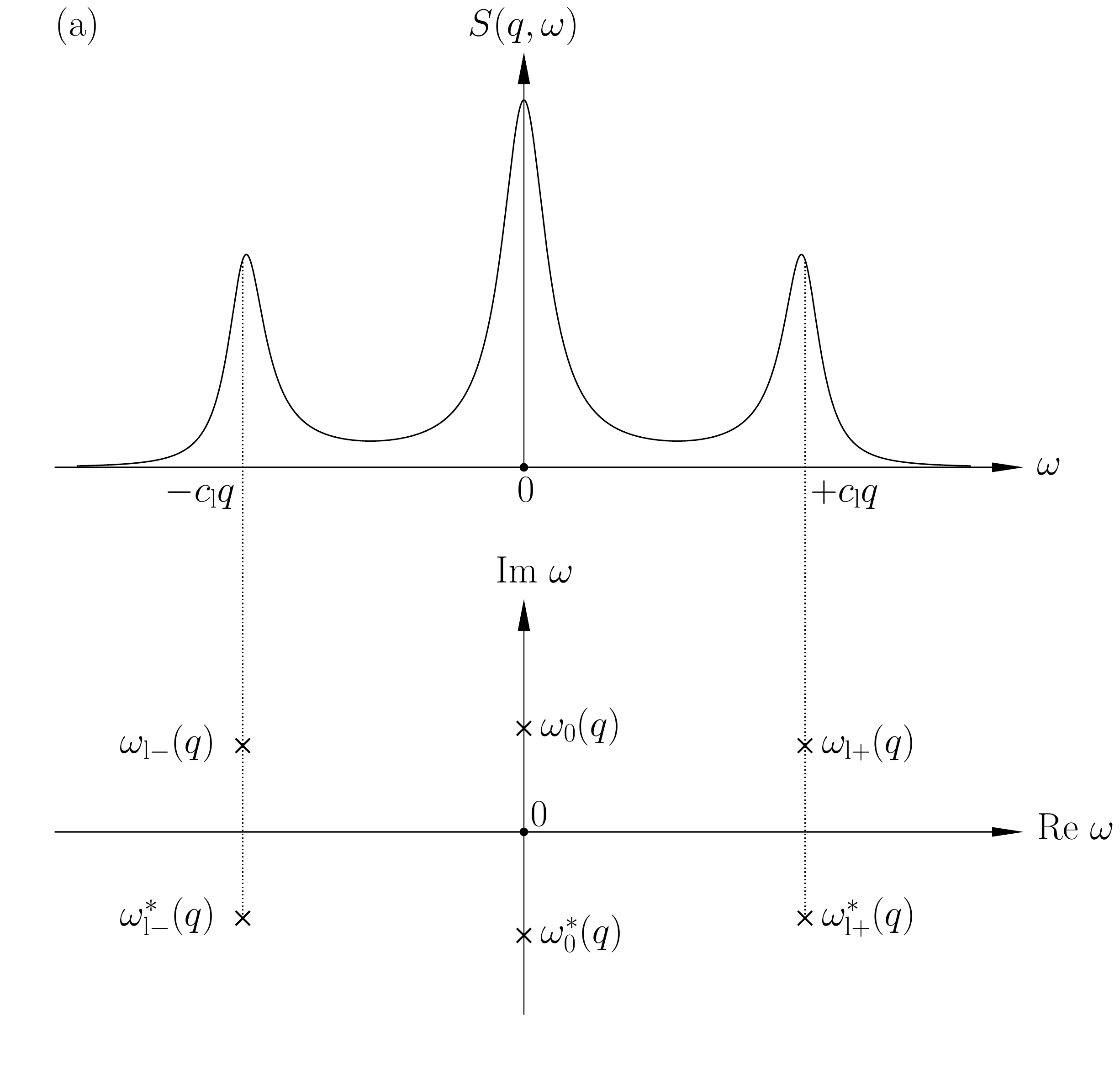}
    \end{minipage}\hfill
    \begin{minipage}{0.5\textwidth}
        \centering
        \includegraphics[width=1.\textwidth]{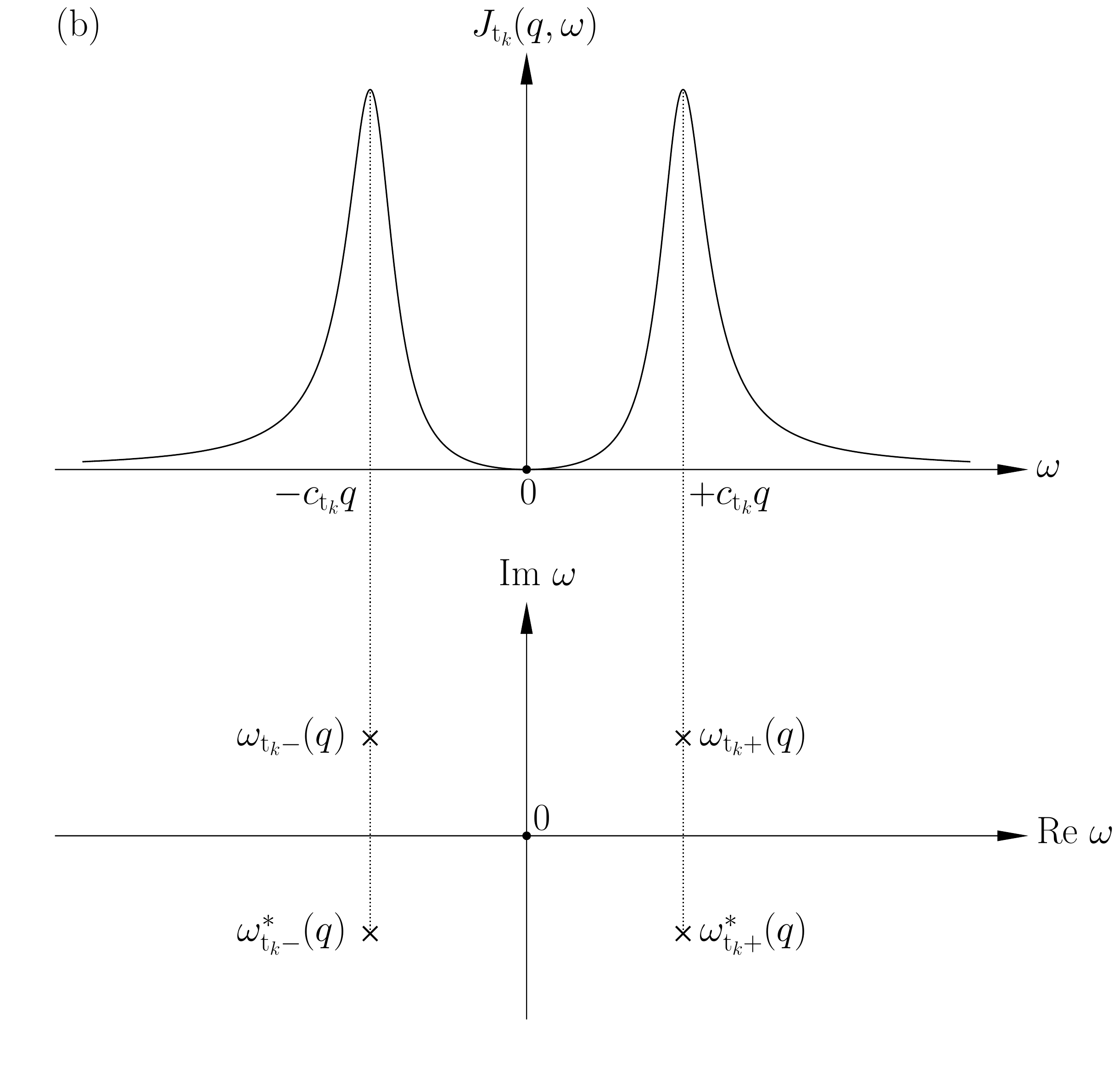}
    \end{minipage}    
\caption[] {Schematic representations of (a) the dynamic structure factor $S(q,\omega)$ given by equation~\eqref{eq:DSF_Full} and (b) the spectral function for the transverse momentum density fluctuations $J_{{\rm t}_k}(q,\omega)$ given by equation~\eqref{eq:JtoCt}.  These spectral functions are plotted versus frequency $\omega$.  Below, their underlying poles are depicted in the plane of complex frequencies for a given value of the wave number~$q=\Vert{\bf q}\Vert$.  The seven hydrodynamic modes of the perfect crystals correspond to the seven resonances given by the three peaks of panel~(a) and the four peaks shown in panel~(b) for $k=1$ and its duplicate for $k=2$.  We note that the presence of  nearby peaks implies that the poles are not always exactly located below the maximum of a peak. This shift is mostly seen for the Brillouin doublet in the panel~(a), and is mainly due to the central peak.}\label{Fig:SFPoles}
\end{figure}

The  dynamic structure factor is obtained from the relation $S({\bf q},\omega)  = 2\, {\rm Re}\, \tilde F({\bf{q}},z={\rm i} \omega)$ in terms of the Laplace transform $\tilde F({\bf{q}},z)$ of the intermediate scattering function. This relation, derived in appendix~\ref{app:FLt}, holds since $F({\bf{q}},t)$ is real and an even function of  time. If the wave vector $\bf q$ is oriented in the directions of table~\ref{Tab:LTCoeffs}, the result, calculated in appendix~\ref{app:calc_tlc},  can be expressed with the notation $\det \boldsymbol{\mathsf M}({q},z)=D_1(\omega)+{\rm i}\, D_2(\omega)$ as
\begin{align}
\frac{S({q},\omega)}{S({q})}&=2\, \frac{N_1(\omega)D_1(\omega)+N_2(\omega)D_2(\omega)}{D^2_1(\omega)+D^2_2(\omega)} ,
\label{eq:DSF_Full}
\end{align}
where
\begin{align}
N_1(\omega)&\equiv -\omega^2+\frac{\left(\gamma-1\right)B_T}{\rho}q^2+\gamma D_T D_v q^4\, , \label{N1}\\
N_2(\omega)&\equiv\omega (D_v+\gamma D_T)q^2\, , \label{N2}\\
D_1(\omega)&\equiv -\omega^2(D_v+\gamma D_T)q^2+\frac{\gamma B^T_{\rm l}D_T}{\rho}q^4\, , \label{D1}\\
D_2(\omega)&\equiv\omega\left[-\omega^2+\frac{B^T_{\rm l}}{\rho}q^2+\frac{\left(\gamma-1\right)B_T}{\rho}q^2+\gamma D_T D_v q^4\right] , \label{D2}
\end{align}
and $S(q)=F(q,0)$.  This dynamic structure factor is depicted in the panel~(a) of figure~\ref{Fig:SFPoles}. 

The poles of the dynamic structure factor are obtained from the roots of the denominator of equation~\eqref{eq:DSF_Full}.  Accordingly, we find the dispersion relations of the modes at leading orders in $q$ as
\begin{align}
\omega_0(q) &= {\rm i}\, \chi \, q^2+\cdots \, , & &\omega_{\rm l\pm}(q) = \pm  c_{\rm l}\, q + {\rm i}\, \Gamma_{\rm l}\, q^2+\cdots \, ,\label{eq:DRl}
\end{align} 
where the dots denote terms vanishing faster than $q^2$ for $q\to 0$, and their complex conjugates $\omega_0^*(q)$ and $\omega^*_{\rm l \pm}(q)$. The coefficient $\chi$ is related to the thermal diffusivity $D_T$, $c_{\rm l }$ is the speed of longitudinal sound waves, and $\Gamma_{\rm l}$ is their acoustic attenuation coefficient.  They are respectively given by 
\begin{align}
\chi & \equiv \frac{\gamma D_T}{1+(\gamma-1)\frac{B_T}{B^T_{\rm l}}}\,,\label{eq:thdi}\\
c_{\rm l } &\equiv \sqrt{\frac{B^T_{\rm l}+(\gamma-1)B_T}{\rho}}\,,\label{eq:cl}\\
\Gamma_{\rm l} &\equiv\frac{1}{2}\left( D_v  + \frac{\gamma D_T}{1+\frac{1}{\gamma-1}\frac{B^T_{\rm l}}{B_T}} \right) .\label{eq:Gaml}
\end{align}
The poles of the dynamic structure factor are shown in the panel~(a) of figure~\ref{Fig:SFPoles}. We observe that the poles are not always exactly located below the maximum of a peak. This shift stems from to the presence of nearby peaks, creating an asymmetry, and here affects the Brillouin doublet.

\begin{figure}[h!] \centering
    \begin{minipage}{0.5\textwidth}
        \centering
        \includegraphics[width=1.\textwidth]{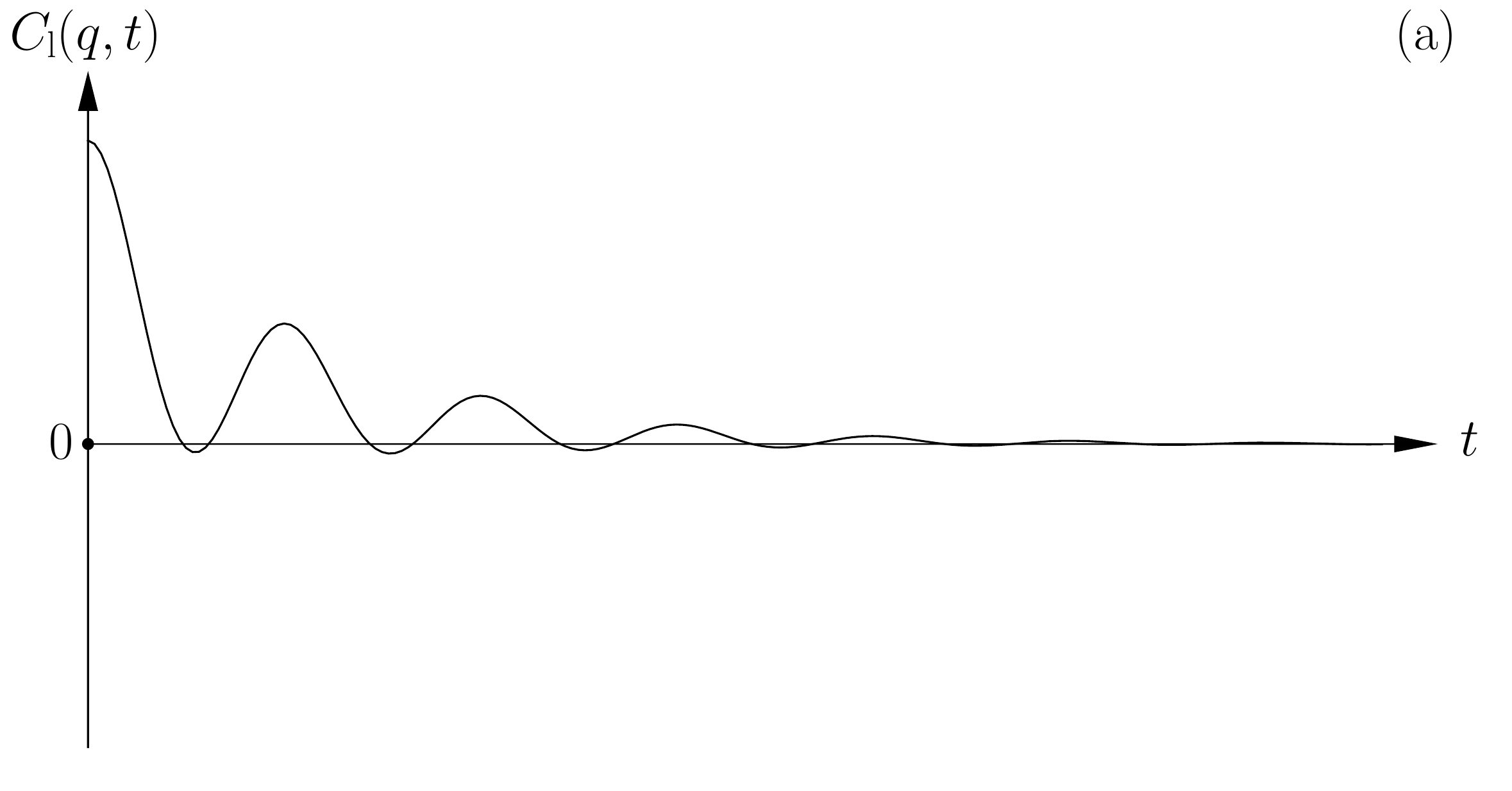}
     \end{minipage}\hfill
    \begin{minipage}{0.5\textwidth}
        \centering
       \includegraphics[width=1.\textwidth]{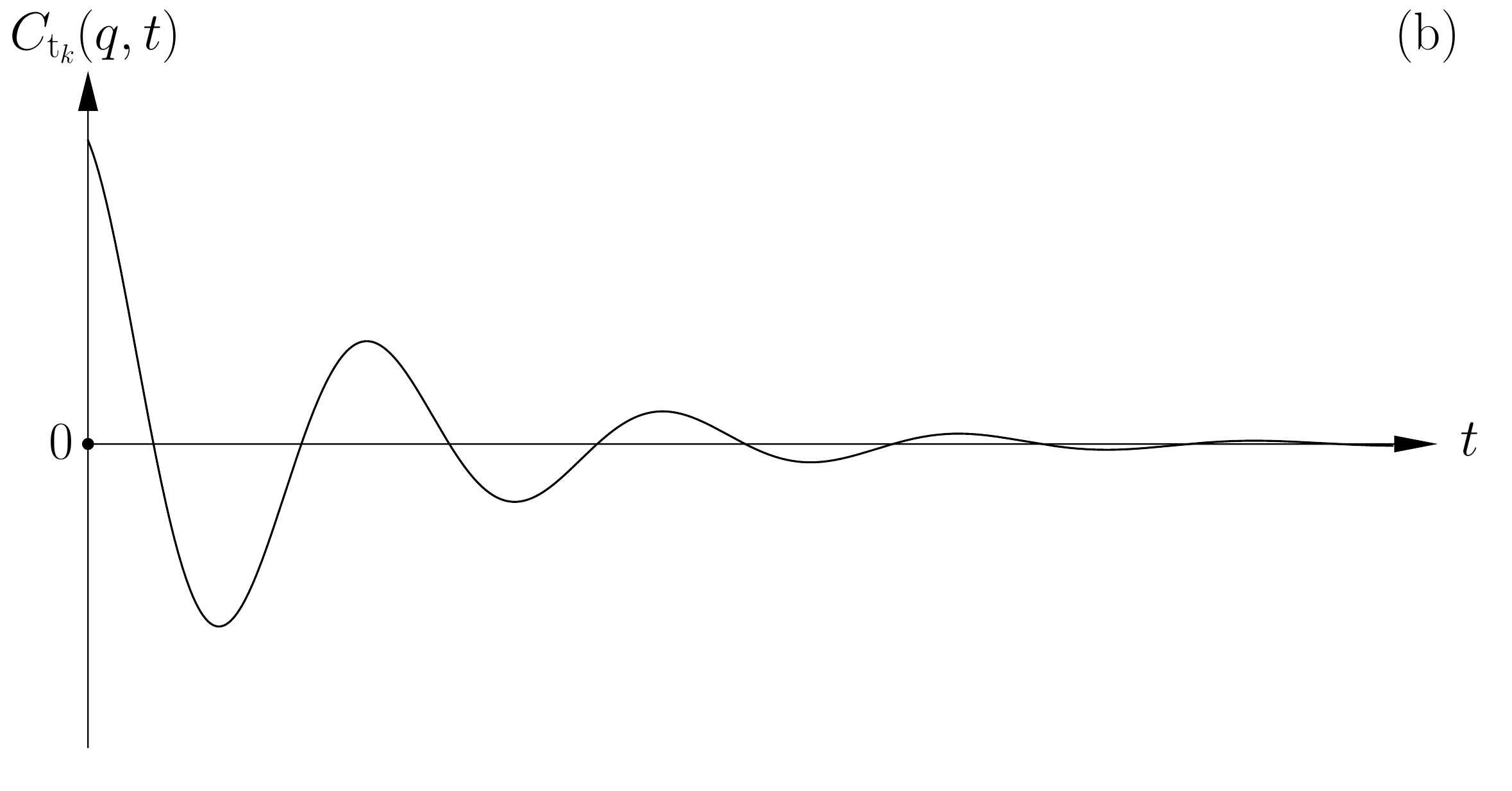}
    \end{minipage}
\caption[] {Schematic representations of (a) the intermediate scattering function $F(q,t)$ given by equation~\eqref{eq:ISF} and (b) the correlation functions for the transverse momentum density fluctuations $C_{{\rm t}_k}(q,t)$ given by equation~\eqref{eq:Ct}.  These correlation functions are plotted versus time $t$ for a given value of the wave number~$q=\Vert{\bf q}\Vert$.}\label{Fig:CF}
\end{figure}

The correlation and spectral functions are further calculated by using a small-$q$ expansion, as carried out in appendix~\ref{app:calc_tlc}.  The intermediate scattering function~\eqref{eq:ISF-mass} is obtained from the inverse Laplace transform of equation~\eqref{eq:Sks} in the limit of small $q$ as
\begin{align}
\frac{F({q},t)}{S({q})} &=\frac{1}{1+(\gamma-1)\frac{B_T}{B^T_{\rm l}}}\left\{(\gamma-1)\frac{B_T}{B^T_{\rm l}} {\rm e}^{-\chi q^2|t|}+\left[\cos(c_{\rm l}q|t|)+\frac{ 3\Gamma_{\rm l}-D_v}{c_{\rm l}}q\sin( c_{\rm l}q|t|)\right]{\rm e}^{-\Gamma_{\rm l} q^2|t|}\right\}
\label{eq:ISF}
\end{align}
for the directions of table~\ref{Tab:LTCoeffs}.  This function has the properties~\eqref{eq:C:eq-station}-\eqref{eq:C:eq-station+microrev} implied by equilibrium stationarity and microreversibility.  Such an intermediate scattering function is schematically depicted in the panel~(a) of figure~\ref{Fig:CF}.

Taking the Fourier transform from time to frequency, the following analytic form is deduced for the corresponding dynamic structure factor,
\begin{align}
\frac{S({q},\omega)}{S({q})} & = \frac{1}{1+(\gamma-1)\frac{B_T}{B^T_{\rm l}}}\left\{(\gamma-1)\frac{B_T}{B^T_{\rm l}} \frac{2\, \chi q^2}{\omega^2+(\chi q^2)^2}+\frac{\Gamma_{\rm l} q^2}{(\omega+c_{\rm l}q)^2+(\Gamma_{\rm l} q^2)^2}+\frac{\Gamma_{\rm l} q^2}{(\omega-c_{\rm l}q)^2+(\Gamma_{\rm l} q^2)^2}\right. \notag\\
&\left. \qquad +\frac{3\Gamma_{\rm l}-D_v}{c_{\rm l}}q\left[\frac{\omega+c_{\rm l}q}{(\omega+c_{\rm l}q)^2+(\Gamma_{\rm l} q^2)^2}-\frac{\omega-c_{\rm l}q}{(\omega-c_{\rm l}q)^2+(\Gamma_{\rm l} q^2)^2}\right]\right\} .\label{eq:DSF_RBP}
\end{align}
The first term in the  bracket of equation~\eqref{eq:DSF_RBP} is a Lorentzian function centered at the origin. It corresponds to a Rayleigh central peak and is associated with the heat mode. This mode is purely dissipative and it has the dispersion relation $\omega_0$ of equation~\eqref{eq:DRl}. The width of the peak is proportional to the coefficient $\chi$, which is related to the thermal diffusivity.  The next terms in the bracket of equation~\eqref{eq:DSF_RBP} form a pair of Lorentzian functions centered at $\pm c_{\rm l}q$. They correspond to a Brillouin doublet and are associated with the pair of longitudinal sound waves propagating with the speed $\pm c_{\rm l}$. They have the dispersion relations $\omega_{\rm l\pm}$ of equation~\eqref{eq:DRl}.  The widths of the peaks are proportional to the coefficient $\Gamma_{\rm l}$, which determines the damping of the sound waves. The symmetries~\eqref{eq:S:eq-station}-\eqref{eq:S:eq-station+microrev} of equilibrium stationarity and microreversibility are satisfied by the function~\eqref{eq:DSF_RBP}.  

%%%%%%%%%%%%%%%%%%%%%%%%%%%%%%%%%%%%%%%%%%%%%%%%%%%%%%
%%%%%%%%%%%%%%%%%%%%%%%%%%%%%%%%%%%%%%%%%%%%%%%%%%%%%%
%%%%%%%%%%%%%%%%%%%%%%%%%%%%%%%%%%%%%%%%%%%%%%%%%%%%%%

\subsection{Transverse correlation and spectral functions} 
\label{sec:transCSF}
%%%%%%%%%%%%%%%%%%%%%%%%%%%%%%%%%%%%%%%%%%%%%%%%%%%%%%

For the directions of table~\ref{Tab:LTCoeffs}, the Laplace transforms of the correlation functions for the transverse components of the velocity and displacement fields are obtained from the sets of transverse linearized equations~\eqref{eq:zv_t}-\eqref{eq:zu_t} as
\begin{align}
\frac{\left\langle\delta \hat{\tilde  v}_{{\rm t}_k}({\bf q},z)\,\delta \hat{v}^*_{{\rm t}_k}({\bf q},0) \right\rangle_{\rm eq}}{\left\langle\delta \hat{v}_{{\rm t}_k}({\bf q},0)\,\delta \hat{v}^*_{{\rm t}_k}({\bf q},0) \right\rangle_{\rm eq}}&=\frac{z}{z\left(z+ \eta_{{\rm t}_k}q^2/\rho\right)+ B^T_{{\rm t}_k}q^2/\rho}\,,\label{eq:vvwovvt}\\
\frac{\left\langle\delta \hat{\tilde  u}_{{\rm t}_k}({\bf q},z)\,\delta \hat{u}^*_{{\rm t}_k}({\bf q},0) \right\rangle_{\rm eq}}{\left\langle\delta \hat{u}_{{\rm t}_k}({\bf q},0)\,\delta \hat{u}^*_{{\rm t}_k}({\bf q},0) \right\rangle_{\rm eq}}&=\frac{z+ \eta_{{\rm t}_k}q^2/\rho}{z\left(z+ \eta_{{\rm t}_k}q^2/\rho\right)+ B^T_{{\rm t}_k}q^2/\rho}\,,\label{eq:uuwouut}
\end{align}
for $k=1,2$ in the two transverse directions, as calculated in appendix~\ref{app:calc_tlc} using the hypothesis of the regression of fluctuations~\cite{MG23}.

The spectral functions of the transverse momentum  density fluctuations are obtained from the relations $J_{{\rm t}_k}({\bf q},\omega) =2\, {\rm Re}\,  \tilde{C}_{{\rm t}_k}({\bf q},z={\rm i}\omega)$ in terms of the Laplace transform $\tilde{C}_{{\rm t}_k}({\bf q},z)$ of the correlation functions~\eqref{eq:v_cf} with $\sigma={\rm t}_k$ for $k=1,2$.  For the directions of table~\ref{Tab:LTCoeffs}, they are thus given by
\begin{align}
\frac{J_{{\rm t}_k}({q},\omega)}{C_{{\rm t}_k}({q},0)} & = \frac{2\, \eta_{{\rm t}_k}q^2\omega^2/\rho}{\left(\omega^2-B^T_{{\rm t}_k}q^2/\rho\right)^2+\left(\eta_{{\rm t}_k}q^2\omega/\rho\right)^2} \, .
\label{eq:JtoCt}
\end{align}
This function has the symmetries~\eqref{eq:S:eq-station}-\eqref{eq:S:eq-station+microrev} of equilibrium stationarity and microreversibility.  

In the hydrodynamic limit, the spectral functions of the transverse momentum  density fluctuations $J_{{\rm t}_k}({q},\omega)$ have poles located at the complex frequencies
\begin{align}
\omega_{{\rm t}_k\pm}(q) & = \pm c_{{\rm t}_k}q+ {\rm i}\, \Gamma_{{\rm t}_k}q^2 + \cdots
\qquad\mbox{with}\qquad
k=1,2 \, ,\label{eq:poles_Jtqw}
\end{align}
and their complex conjugates $\omega_{{\rm t}_k\pm}^*(q)$, where the speeds of the transverse sound waves and their acoustic attenuation coefficients are respectively given by
\begin{align}
c_{{\rm t}_k}\equiv \sqrt{\frac{B^T_{{\rm t}_k}}{\rho}} 
\qquad\mbox{and}\qquad
\Gamma_{{\rm t}_k}\equiv\frac{\eta_{{\rm t}_k}}{2\rho} \, .
\label{eq:c_t+G_t}
\end{align}
An example of spectral function characterizing the transverse momentum density fluctuations is depicted in the panel~(b) of figure~\ref{Fig:SFPoles}, together with the underlying poles. The two spectral functions $J_{{\rm t}_k}({q},\omega)$ have two peaks each, located at $\pm  c_{{\rm t}_k}q$, which correspond to two pairs of transverse sound  waves propagating with the speeds $\pm c_{{\rm t}_k}$. The widths of the peaks are proportional to the acoustic attenuation coefficients $\Gamma_{{\rm t}_k}$, which determine the damping of the transverse sound waves.

For the directions of table~\ref{Tab:LTCoeffs}, the correlation functions of the transverse momentum density fluctuations are obtained from the Fourier transform of equation~\eqref{eq:JtoCt} from frequency to time as 
\begin{align}
\frac{C_{{\rm t}_k}({q},t)}{C_{{\rm t}_k}({q},0)} &= {\rm e}^{-\Gamma_{{\rm t}_k}q^2|t|}\left[\cos\left(q|t|\sqrt{c_{{\rm t}_k}^2-\Gamma_{{\rm t}_k}^2q^2}\right)-\frac{\Gamma_{{\rm t}_k}q}{\sqrt{c_{{\rm t}_k}^2-\Gamma_{{\rm t}_k}^2q^2}}\sin\left(q|t|\sqrt{c_{{\rm t}_k}^2-\Gamma_{{\rm t}_k}^2q^2}\right)\right]  
\label{eq:Ct}
\end{align}
for $k=1,2$. This function satisfies the properties~\eqref{eq:C:eq-station}-\eqref{eq:C:eq-station+microrev} due to  equilibrium stationarity and microreversibility.  Such a correlation function is depicted in the panel~(b) of figure~\ref{Fig:CF}.

%%%%%%%%%%%%%%%%%%%%%%%%%%%%%%%%%%%%%%%%%%%%%%%%%%%%%%

\subsection{Comparison between the crystal and the fluid} 
\label{sec:crystal-fluid}
%%%%%%%%%%%%%%%%%%%%%%%%%%%%%%%%%%%%%%%%%%%%%%%%%%%%%%

Here above, we have shown that the seven hydrodynamic modes of the perfect crystal can be identified from the seven resonance peaks of the spectral functions~\eqref{eq:DSF_Full} and~\eqref{eq:JtoCt} for mass and transverse momentum densities, respectively.  In the crystal, the characteristic feature of the hydrodynamic spectrum is the presence of four transverse sound modes with the speeds $\pm c_{{\rm t}_1}$ and $\pm c_{{\rm t}_2}$, in addition to the two longitudinal sound modes with the speeds $\pm c_{\rm l}$.  The transverse sound modes arise from the anisotropy and spatial periodicity of the crystal, which is generated by the spontaneous symmetry breaking of spatial translations.

In contrast, the fluid is isotropic and uniform and, moreover, its shear modulus is equal to zero, $B_{44}^T=0$.
Alone, isotropy implies that
\begin{align}
B_{11}^T - B_{12}^T = 2 \, B_{44}^T
\qquad\mbox{and}\qquad
\eta_{11} - \eta_{12} = 2 \, \eta_{44} \, ,
\end{align}
as in amorphous solids.  However, fluids are flowing because they cannot resist a shear stress, which is expressed by the further condition that $B_{44}^T=0$.  As a consequence, we have in fluids that $B_{11}^T=B_{12}^T$ and the isothermal longitudinal stress-strain coefficient becomes equal to the isothermal bulk modulus $B_T$, while the transverse stress-strain coefficients are equal to zero,
\begin{align}
B_{\rm l}^T = B_T
\qquad\mbox{and}\qquad
B_{{\rm t}_1}^T =  B_{{\rm t}_2}^T =  0 \, ,
\end{align}
whereupon the speeds of the transverse sound waves are also equal to zero, $c_{{\rm t}_1} =  c_{{\rm t}_2} =  0$, in all the directions of table~\ref{Tab:LTCoeffs}.
Therefore, for the fluid, the two peaks of the spectral function~\eqref{eq:JtoCt} for the transverse momentum density fluctuations seen in the panel (b) of figure~\ref{Fig:SFPoles} merge and they become the zero-frequency peak associated with the diffusive shear modes existing in fluids~\cite{MG23}.  Consequently, the seven hydrodynamic modes of  the perfect crystal reduce to the five hydrodynamic modes of the fluid.

In fluids, the three viscosity coefficients of cubic crystals reduce to the two shear $\eta$ and bulk $\zeta$ viscosities of fluids according to
\begin{align}
\eta_{11} = \zeta + \frac{4}{3} \, \eta \, , \qquad
\eta_{12} = \zeta - \frac{2}{3} \, \eta \, , \qquad
\mbox{and}\qquad
\eta_{44} = \eta \, ,
\end{align}
so that the isotropy condition $\eta_{11} - \eta_{12} = 2 \, \eta_{44}$ is satisfied, but the shear viscosity $\eta$ remains positive.  Accordingly, the longitudinal and transverse viscosities of table~\ref{Tab:LTCoeffs} become
\begin{align}
\eta_{\rm l} = \eta_{11} = \zeta + \frac{4}{3} \, \eta
\qquad\mbox{and}\qquad
\eta_{{\rm t}_1} =  \eta_{{\rm t}_2} = \eta_{44} = \eta \, .
\end{align}

For these reasons, the correlation and spectral functions here calculated for the perfect cubic crystal reduce to those that are known for the fluid \cite{MG23,BP76,BY80}.  In particular, the dynamic structure factor~\eqref{eq:DSF_RBP} of the crystal reduces to the one of the fluid given by the formula~(45) in reference~\cite{MG23}, because the speed~\eqref{eq:cl} of the longitudinal sound waves becomes equal to $c_{\rm l}=\sqrt{\gamma B_T/\rho}=c_s$, which is the speed of sound in fluids, and similarly for the acoustic attenuation coefficient~\eqref{eq:Gaml}, which becomes $\Gamma_{\rm l}=\Gamma=[D_v+D_T(\gamma-1)]/2$ with $D_v=(\zeta+\frac{4}{3}\eta)/\rho$ in fluids, and the coefficient~\eqref{eq:thdi}, which reduces to the thermal diffusivity $\chi=D_T$ in agreement with equation~(47) of reference~\cite{MG23} for fluids.  Moreover, the spectral function~\eqref{eq:JtoCt} for the transverse momentum density fluctuations in the perfect cubic crystal reduces to the formula~(46) of reference~\cite{MG23} for the momentum density fluctuations in the fluid, since the speeds of the transverse sound waves are therein vanishing, $c_{{\rm t}_k}=\sqrt{B_{{\rm t}_k}^T/\rho}=0$, and the transverse viscosities of the crystal $\eta_{{\rm t}_k}$ become equal to the shear viscosity $\eta$ of the fluid.

Therefore, the results here obtained for perfect cubic crystals are consistent with those deduced for fluids in references~\cite{MG23,BP76,BY80}.  The comparison shows that the seven hydrodynamic modes of perfect crystals arise from the emergence of a shear modulus $B_{44}^T\ne 0$ due to the spontaneous symmetry breaking of spatial translations and the resulting long-range order in the crystalline phase.

%%%%%%%%%%%%%%%%%%%%%%%%%%%%%%%%%%%%%%%%%%%%%%%%%

\subsection{Dispersion relations}
\label{subsec:DR}

From the poles of the spectral functions $S({q},\omega)$ in equation~\eqref{eq:DSF_Full} and $J_{{\rm t}_k}({q},\omega)$ in equation~\eqref{eq:JtoCt}, we have obtained the dispersion relations~\eqref{eq:DRl} and~\eqref{eq:poles_Jtqw}, respectively, and identified the seven hydrodynamic modes of the perfect crystal.  In order to compare the theoretical predictions given by the spectral functions~\eqref{eq:DSF_Full} and~\eqref{eq:JtoCt} with their numerical calculations using molecular dynamics simulations, we need to evaluate the coefficients $\chi$, $c_{\rm l}$, $\Gamma_{\rm l}$, $c_{{\rm t}_k}$, and $\Gamma_{{\rm t}_k}$ given by equations~
\eqref{eq:thdi}, \eqref{eq:cl}, \eqref{eq:Gaml}, and~\eqref{eq:c_t+G_t}.  They are expressed in terms of the equilibrium and nonequilibrium hydrodynamic properties of the crystal, which have been computed by numerically simulating the molecular dynamics of the hard-sphere system and the method of Helfand moments in reference~\cite{MG23_primo}.  From the so-computed data for the thermodynamic, elastic, and transport properties, we can obtain the coefficients appearing in the dispersion relations as function of the density $n_*$. The results are reported in tables~\ref{Tab:PC100},~\ref{Tab:PC110}, and~\ref{Tab:PC111} for ${\bf q}$ in the directions $[100]$, $[110]$, and $[111]$, respectively, and they are shown in figure~\ref{Fig:PC}. 

The speeds of longitudinal and transverse sound waves, and their attenuation coefficients versus the density for the three directions are shown in figure~\ref{Fig:cs}. The speed is always larger for the longitudinal  than the transverse sound waves. The same observation holds for the attenuation coefficients. Moreover, the speed of the second transverse sound mode in the direction $[110]$ is the same as the speed of the two transverse sound modes in the direction [100], as expected from the cubic symmetry.  The speeds and the acoustic attenuation coefficients of the sound waves diverge as $(\sqrt{2}-n_*)^{-1}$ near the close-packing density $n_*=\sqrt{2}$. This scaling is consistent with the divergence of the collision frequency~\cite{MG23_primo}.

The dispersion relations for the hard-sphere crystal at densities $n_*=1.037$ and $n_*=1.3$ for the directions $[100]$, $[110]$, and $[111]$ are depicted in figures~\ref{Fig:DR_1d037} and~\ref{Fig:DR_1d3}.  As already noted, the speed of longitudinal sound waves is larger than those of transverse sound waves. As expected, the heat mode is not propagating. The  speeds of the two transverse sound waves and their attenuation coefficients  take identical values in each one of the directions $[100]$ and $[111]$, which can be seen from table~\ref{Tab:LTCoeffs}.  

Next, the coefficients in tables~\ref{Tab:PC100},~\ref{Tab:PC110}, and~\ref{Tab:PC111} are used to plot the spectral functions~\eqref{eq:DSF_Full} and~\eqref{eq:JtoCt} predicted by hydrodynamics in comparison with those computed using molecular dynamics simulations, as explained in the following section~\ref{sec:MDHS}.

%%%%%%%%%%%%%%%%%%%%%%%%%%%%%%%%%%%%%%%%%%%%%%%%%
\section{Correlation and spectral functions for the perfect hard-sphere crystal}
%\section{Correlation and spectral functions by molecular dynamics}
\label{sec:MDHS}

In order to test the predictions of the hydrodynamics of perfect crystals for the correlation and spectral functions, we simulate the dynamics of the hard-sphere crystal using the event-driven algorithm already presented in references~\cite{MG23_primo,MG23} and we compute the intermediate scattering function~\eqref{eq:ISFMD} and the momentum density correlation functions~\eqref{eq:mom_cf} from the molecular dynamics simulation.  A Fourier transform from time to frequency gives the dynamic structure factor~\eqref{eq:DSF-dfn} and the spectral functions~\eqref{eq:spctr_fns-dfn} of momentum density fluctuations.

\subsection{The hard-sphere dynamics}

The hard-sphere system is composed of $N$ identical particles of mass $m$ and diameter $d$ moving in a cubic domain of sides $L$ and volume $V=L^3$ with periodic boundary conditions.  The edges of this cubic domain are chosen along the $x$, $y$, and $z$ axes.  The event-driven algorithm simulates the motion of the particles as free flights interrupted by binary elastic collisions~\cite{H97}.  The simulation is performed in the $(N,V,E)$-ensemble and the total momentum is set to ${\bf P}=0$.  The equilibrium temperature is thus set equal to $k_{\rm B}T=(2/3)(E/N)$.

The hard spheres are initially located on a  fcc lattice composed of $M$ cubic cells of size $a$ in each direction, such that $L=Ma$.  Each cell contains four particles initially at the positions ${\bf R}_j =x_j {\bf e}_x +y_j {\bf e}_y + z_j {\bf e}_z$ with $j\in\{1,2,3,4\}$ and $a^{-1}(x_j,y_j,z_j)\in\{(\frac{1}{4},\frac{1}{4},\frac{1}{4}),(\frac{3}{4},\frac{3}{4},\frac{1}{4}),(\frac{3}{4},\frac{1}{4},\frac{3}{4}),(\frac{1}{4},\frac{3}{4},\frac{3}{4})\}$ \cite{AM76}.  Therefore, all the sites of the fcc lattice are occupied with a hard sphere, as required to simulate a perfect crystal.  This configuration allows us to reach values for the particle density up to the close-packing density to simulate the system in the crystalline phase.  The spatially averaged mean particle density is thus equal to $n=4/a^3=N/L^3$, so that the size of the cubic simulation domain should be taken as $L=M(4/n)^{1/3}$ for a fixed value $n$ of the particle density.  

The initial momenta of the hard spheres are randomly distributed with the constraint that their total momentum is equal to zero.  The event-driven algorithm generates the trajectories of the hard spheres.  Their positions ${\bf r}_i(t)$ and momenta ${\bf p}_i(t)$ are thus known at any time $t$ of the simulation.  First, the dynamics is run during some transient time $t_{\rm transient}$ in order to reach statistical equilibration before collecting data.  
Next, the trajectories are sampled into $n_{\rm steps}$ discrete time steps~$\Delta t$.  The equilibrium statistical average of any quantity $X$ is evaluated as $\langle X\rangle_{\rm eq} = N_{\rm stat}^{-1} \sum_{k=1}^{N_{\rm stat}} X^{(k)}$ with a large enough number $N_{\rm stat}$ of trajectories.  In this way, the correlation and spectral functions are computed by statistics over trajectories forming the $(N,V,E)$-ensemble with zero total momentum.

Because of the periodic boundary conditions, the wave vector takes the discrete values ${\bf q} = (2\pi/L)\left(n_x{\bf e}_x+n_y{\bf e}_y+n_z{\bf e}_z \right)$. The integer values $(n_x,n_y,n_z)=(1,0,0)$, $(n_x,n_y,n_z)=(1,1,0)$, and $(n_x,n_y,n_z)=(1,1,1)$ are considered in order for the wave vector to be oriented in the special directions of table~\ref{Tab:LTCoeffs} and to reach the hydrodynamic regime.

The quantities of interest are rescaled using the mass $m$ and the diameter $d$ of the hard spheres and the temperature $k_{\rm B}T$, which take the unit value in the simulations.  Accordingly, the results are presented in terms of dimensionless quantities denoted with an asterisk as subscript.  The particle density, wave number, frequency, sound speeds, diffusivities, and spectral functions are respectively given in terms of the corresponding dimensionless quantities by
\begin{align}
& n = \frac{n_*}{d^3} \, , \qquad q = \frac{q_*}{d}\, , \qquad \omega = \frac{\omega_*}{d}\, \sqrt{\frac{k_{\rm B}T}{m}} \, , \\
& c_\sigma = c_{\sigma *} \sqrt{\frac{k_{\rm B}T}{m}} \, , \qquad 
\Gamma_\sigma= \Gamma_{\sigma *} \, d \, \sqrt{\frac{k_{\rm B}T}{m}} \, , \qquad
\chi = \chi_* \, d \, \sqrt{\frac{k_{\rm B}T}{m}} \, , \\
& \frac{S(q,\omega)}{S(q)} = \left[ \frac{S(q,\omega)}{S(q)} \right]_* \, d \, \sqrt{\frac{m}{k_{\rm B}T}} \, ,
\qquad\mbox{and}\qquad
\frac{J_\sigma(q,\omega)}{C_\sigma(q,0)} = \left[ \frac{J_\sigma(q,\omega)}{C_\sigma(q,0)} \right]_* \, d \, \sqrt{\frac{m}{k_{\rm B}T}}
\end{align}
for $\sigma\in\{{\rm l},{\rm t}_1,{\rm t}_2\}$.

\subsection{Results for the correlation and spectral functions} 

The intermediate scattering function $F({\bf q},t)$ and the momentum  density correlation functions $C_\sigma({\bf q},t)$  with $\sigma\in\{{\rm l},{\rm t}_1,{\rm t}_2\}$ are computed using equations~\eqref{eq:ISFMD} and~\eqref{eq:mom_cf}, respectively, with the molecular dynamics simulation of a system of $N=2048$ hard spheres, corresponding to $M=8$, in the crystalline phase for the densities $n_*=1.037$ and $n_*=1.3$.  After the transient time $t_{\rm * transient}=50$, statistics is carried out over $N_{\rm stat}=10^4$ trajectories sampled at discrete time steps $\Delta t_*=0.01$.  The number of steps $n_{\rm steps}$ varies for the directions and the densities considered.  The smallest possible value is used for the wave number $q=\Vert{\bf q}\Vert$ associated with the wave vector ${\bf q}$ in the directions $[100]$, $[110]$, and $[111]$.   The corresponding spectral functions $S({\bf q},\omega)$ and $J_{\sigma}({\bf q},\omega)$ are obtained by numerical Fourier transform. All these functions are normalized by the value of the corresponding correlation function at time $t=0$.

The correlation and spectral functions obtained from the simulation are compared to the analytical expressions predicted by hydrodynamics and given by equation~\eqref{eq:ISF} for $F({\bf q},t)$, equation~\eqref{eq:DSF_Full} for $S({\bf q},\omega)$, equation~\eqref{eq:Ct} for $C_{{\rm t}_k}({\bf q},t)$, and equation~\eqref{eq:JtoCt} for $J_{{\rm t}_k}({\bf q},\omega)$, using the coefficients given in tables~\ref{Tab:PC100}-\ref{Tab:PC111} with data from reference~\cite{MG23_primo}.  The analytical expressions for $C_{\rm l}({\bf q},t)$ and $J_{\rm l}({\bf q},\omega)$ are obtained from the intermediate scattering function and the dynamic structure factor with equations~\eqref{eq:lca_hydro} and~\eqref{eq:lsa_hydro}. 

The results are presented in figures~\ref{Fig:CF100-1.037}-\ref{Fig:CF111-1.3}, showing an excellent agreement between the numerical functions and those predicted by the hydrodynamics of the perfect crystal.  In these figures, the intermediate scattering functions $F(q,t)$ present damped oscillations caused by the longitudinal sound waves that are superposed onto the exponential decay due to the heat mode, as expected from equation~\eqref{eq:ISF} and the panel~(a) of figure~\ref{Fig:CF}.  Accordingly, three resonance peaks appear in the dynamic structure factors $S(q,\omega)$, namely, the central Rayleigh peak of the heat mode, and the Brillouin doublet of the longitudinal sound modes, as in fluids.  As a consequence of equation~\eqref{eq:lsa_hydro}, the central Rayleigh peak has disappeared in the longitudinal momentum density spectral functions $J_{\rm l}({q},\omega)$, which only present the same Brillouin doublet as in the dynamic structure factors, and the longitudinal momentum density correlation functions $C_{\rm l}({q},t)$ have corresponding damped oscillations, also as in fluids.  However, in contrast to fluids, damped oscillations appear in the transverse momentum density correlation functions $C_{{\rm t}_k}({q},t)$, because the diffusive shear modes of the fluid are turned into the transverse sound waves of the crystal after the spontaneous symmetry breaking of the spatial translations.  The behavior observed in the numerical simulations of the functions $C_{{\rm t}_1}({q},t)$ and $C_{{\rm t}_2}({q},t)$ agrees very well with the expectation from equation~\eqref{eq:Ct} and the panel~(b) of figure~\ref{Fig:CF}.  Therefore, the corresponding spectral functions $J_{{\rm t}_k}({q},\omega)$ present two separated resonance peaks at the opposite frequencies $\pm c_{{\rm t}_k}q$ of the transverse sound modes, which are the signature of the crystalline phase.  The excellent agreement between the numerical and the predicted functions supports the validity of the microscopic computations of the hydrodynamic properties obtained in reference~\cite{MG23_primo} for the hard-sphere crystal.  

We note that the oscillations observed in some of the transverse momentum  density correlation functions, such as in figure~\ref{Fig:CF100-1.037}, are spurious and stem from the numerical Fourier transform. These  spurious oscillations occur when the correlation function has not fully decayed over the time interval considered for the Fourier transform. We also note that the  noise that appears in some of the dynamical structure factors, for example in figure~\ref{Fig:CF111-1.037}, is due to long-time fluctuations in the correlation function, which would require larger statistics to be removed.

%%%%%%%%%%%%%%%%%%%%%%%%%%%%%%%%%%%%%%%%%%%%%%%%%
\section{Conclusion and perspectives}
\label{sec:conclusion}

In this paper, we have obtained the time-dependent correlation functions and the corresponding spectral functions for the hydrodynamics of perfect cubic crystals in two different approaches, using the hard-sphere system as the vehicle of our study.  

On the one hand, the correlation and spectral functions characterizing the fluctuations of given wave vector ${\bf q}$ for the mass and momentum densities have been directly computed using molecular dynamics simulations in the hydrodynamic regime by taking the wave number $q=\Vert{\bf q}\Vert$ of the fluctuating Fourier modes to be as small as possible.  The dynamics of the system has been simulated with an event-driven algorithm for a system of $N=2048$ hard spheres at the densities $n_*=1.037$ and $n_*=1.3$ in the crystalline phase, where the lattice is fcc.

On the other hand, these correlation and spectral functions have been calculated using the hydrodynamics of perfect cubic crystals.  Analytical expressions have been deduced for the intermediate scattering function and the dynamic structure factor characterizing the fluctuations of the longitudinal modes, and further correlation and spectral functions for the fluctuations of the transverse modes.  These functions depend on the thermodynamic, elastic, and transport coefficients that we have obtained in our previous paper~\cite{MG23_primo}.  

We observe an excellent agreement between the functions calculated in the two approaches, providing strong evidence for the validity of the microscopic hydrodynamic theory of crystals.  The study we have carried out in this paper shows that the seven hydrodynamic modes of perfect crystals can be identified with the resonance peaks of the spectral functions, i.e., the dynamic structure factor giving the frequency content of the mass density fluctuations and the spectral functions for the two transverse directions of the momentum density fluctuations.  As in fluids, the dynamic structure  factor has a Rayleigh peak at zero frequency caused by the diffusive mode of heat conduction and a Brillouin doublet of peaks due to the two longitudinal sound modes.  However, in contrast to fluids, each one of the two spectral functions for the transverse momentum density fluctuations has two separated peaks at opposite non-zero frequencies.  Hence, the four peaks of these spectral functions correspond to the four transverse sound modes, which form a key feature of crystals.  The comparison between the crystal and the fluid hydrodynamic properties is discussed in subsection~\ref{sec:crystal-fluid}.  The frequencies where the peaks are located are given by the real part of the dispersion relations, ${\rm Re}\,\omega_r=c_r q$, which are proportional to the propagation speed $c_r$ and the wave number $q$ of the sound waves  in the hydrodynamic regime.  The speeds depend on the isothermal stress-strain coefficients, the isothermal bulk modulus, the ratio of heat capacities, and the spatially averaged mass density.  In addition, the widths of the resonance peaks are given by the imaginary part of the dispersion relations going as ${\rm Im}\,\omega_r=\Gamma_r q^2$ in terms of some damping coefficient $\Gamma_r$ and the square $q^2$ of the wave number.  The damping coefficients are the acoustic attenuation coefficients for the sound modes and the diffusivity for the heat mode.  The widths of the peaks are thus determined by the transport coefficients of the crystal.  Since the hard-sphere crystal is cubic, these coefficients include three viscosities and one heat conductivity.  Therefore, the excellent agreement observed between the functions obtained, on the one hand, by molecular dynamics simulations and, on the other hand, by hydrodynamics provides a verification of the values of the transport coefficients computed in our previous paper~\cite{MG23_primo} with Einstein-Helfand formulas.

Furthermore, the results show that the acoustic attenuation coefficients of the sound modes and the diffusivity of the heat mode increase with the particle density, as for the speeds of the sound waves.  For the hard-sphere crystal, these quantities diverge as $(\sqrt{2}-n_*)^{-1}$ near the close-packing density $n_*=\sqrt{2}$, which is consistent with the divergence of the collision frequency~\cite{MG23_primo}.

In the following paper~\cite{MG23_tertio}, we will locate numerically the poles at complex frequencies for the spectral functions computed by molecular dynamics simulations, as a further method to obtain the hydrodynamic coefficients in addition to the method of Helfand moments and to provide an additional comparison for the values of the elastic and transport coefficients in the perfect hard-sphere crystal.

%%%%%%%%%%%%%%%%%%%%%%%%%%%%%%%%%%%%%%%%%%%%%%%%%%

\section*{Acknowledgements}

The authors acknowledge the support of the Universit\'e Libre de Bruxelles (ULB) and the Fonds de la Recherche Scientifique de Belgique (F.R.S.-FNRS) in this research. J.~M. is a Postdoctoral Researcher of the Fonds de la Recherche Scientifique de Belgique (F.R.S.-FNRS).  Computational resources have been provided by the Consortium des Equipements de Calcul Intensif (CECI), funded by the Fonds de la Recherche Scientifique de Belgique (F.R.S.-FNRS) under Grant No. 2.5020.11 and by the Walloon Region.

%%%%%%%%%%%%%%%%%%%%%%%%%%%%%%%%%%%%%%%%%%%%%%%%%
\appendix
%%%%%%%%%%%%%%%%%%%%%%%%%%%%%%%%%%%%%%%%%%%%%%%%%

\section{Fourier-Laplace transform}
\label{app:FLt}

The Fourier-Laplace transform $\tilde{f}({\bf q},z)$ of the function $f({\bf r},t)$ is defined as
\begin{align}
\tilde{f}({\bf q},z)&\equiv\int_0^\infty \text{d}t\ {\rm e}^{-zt}\int_{\mathbb{R}^3} \text{d}{\bf r}\ {\rm e}^{{\rm i}{\bf q}\cdot{ \bf r}}f({\bf r},t) \, .
\end{align}
The inverse transform is given for $t>0$ by
\begin{align}
f({\bf r},t)&=\frac{1}{2\pi{\rm i}}\int_{c-{\rm i}\infty}^{c+{\rm i}\infty} \text{d}z\ {\rm e}^{zt}\int_{\mathbb{R}^3} \frac{\text{d}{\bf q}}{(2\pi)^3}\ {\rm e}^{-{\rm i}{\bf q}\cdot{\bf r}}\tilde{f}({\bf q},z)\, ,\label{eq:invLFT}
\end{align}
where $c$ is a constant larger than the real part ${\rm Re}\, z_r$ of all the singularities of the function $\tilde f({\bf q},z)$. For functions that are well behaved at infinity, i.e., such that $f({\bf r},t)\rightarrow 0$ for $\Vert{\bf r}\Vert\rightarrow \pm \infty$ or $t\rightarrow \infty$, we have 
\begin{align}
\widetilde{\partial_t{f}}({\bf q},z)&=-f({\bf q},0)+z\tilde{f}({\bf q},z)\, , &&
\widetilde{{\boldsymbol \nabla}{f}}({\bf  q},z)=-{\rm i}\,{\bf q}\,\tilde{f}({\bf q},z)\, ,&&
\widetilde{{\nabla}^2{f}}({\bf q},z)=-{q}^2\tilde{f}({\bf q},z)\, , \label{eq:FLpartial_r}
\end{align}
where $q=\Vert{\bf q}\Vert$. 

Moreover, if $f({\bf q},t)=f({\bf q},-t)$ and $f({\bf q},t)\in\mathbb{R}$, we have $f({\bf q},\omega)=2\,{\rm Re}\,\tilde{f}({\bf q},z={\rm i}\omega)$, since
\begin{align}
f({\bf q},\omega) &= \int_{\mathbb{R}}\text{d}t\, {\rm e}^{-{\rm i}\omega t} f({\bf q},t) = \int_{0}^{\infty}\text{d}t \left( {\rm e}^{{\rm i}\omega t}+{\rm e}^{-{\rm i}\omega t}\right) f({\bf q},t) = \tilde{f}({\bf q},z=-{\rm i}\omega) + \tilde{f}({\bf q},z={\rm i}\omega) \notag\\
&= \tilde{f}^*({\bf q},z={\rm i}\omega) + \tilde{f}({\bf q},z={\rm i}\omega) = 2\, {\rm Re}\, \tilde{f}({\bf q},z={\rm i}\omega)\, . \label{eq:FTtoLT}
\end{align}
This formula holds in particular if the function $f({\bf q},t)$ is taken as the correlation function~\eqref{eq:correl-dfn} with $B=A^*$ for $A$ given by the complex conjugate of the spatial Fourier transform~\eqref{eq:n(q)} of the particle density or~\eqref{eq:g(q)} for a component of the momentum density.  In these cases, the property~\eqref{eq:C:microrev} implies that $f({\bf q},t)=f({\bf q},-t)$ and the property~\eqref{eq:C:eq-station+microrev} that $f({\bf q},t)\in\mathbb{R}$, whereupon equation~\eqref{eq:FTtoLT} is satisfied.

%%%%%%%%%%%%%%%%%%%%%%%%%%%%%%%%%%%%%%%%%%%%%%%%%
\section{Fourier-Laplace transforms of the linearized hydrodynamic equations}
\label{app:calc}
%%%%%%%%%%%%%%%%%%%%%%%%%%%%%%%%%%%%%%%%%%%%%%%%%
\subsection{Calculations of the longitudinal and transverse linearized equations}
\label{app:calc_tle}

\paragraph{From equation~\eqref{eq:fl-u} for the strain tensor to the longitudinal and transverse equations~\eqref{eq:zrho_l} and \eqref{eq:zu_t}.} Using the decomposition~\eqref{eq:vlt} of the velocity field ${\bf v}({\bf q})$, we get
\begin{align}
\frac{{\rm i}}{2}\left[q^a \delta v^b(\mathbf{q})+q^b\delta v^a(\mathbf{q})\right]  & =  \frac{{\rm i}}{2}\, q \sum_\sigma \delta v_\sigma({\bf q}) \left(e^a_{\rm l} \, e^b_{\sigma} + e^b_{\rm l} \, e^a_{\sigma}\right) .
\end{align}
Since $\delta u^{ab}(\mathbf{q})$ has the similar decomposition~\eqref{eq:uablt}, equation~\eqref{eq:fl-u} for the  strain tensor becomes
\begin{align}
\sum_\sigma \left[z\, \delta \tilde u_\sigma(\mathbf{q},z)- \delta \tilde v_\sigma(\mathbf{q},z)-\delta u_\sigma(\mathbf{q},0)\right] \left(e^a_{\rm l} \, e^b_{\sigma} + e^b_{\rm l} \, e^a_{\sigma}\right) &= 0 \,,
\end{align}
leading to the longitudinal and transverse equations 
\begin{align}
z\, \delta \tilde u_{\rm l}(\mathbf{q},z)- \delta \tilde v_{\rm l}(\mathbf{q},z)&=\delta u_{\rm l}(\mathbf{q},0)\, , \label{eq:u(q)_l}\\
z\, \delta \tilde u_{{\rm t}_k}(\mathbf{q},z)- \delta \tilde v_{{\rm t}_k}(\mathbf{q},z)&=\delta u_{{\rm t}_k}(\mathbf{q},0)\, . \label{eq:u(q)_t}
\end{align}
The longitudinal equation~\eqref{eq:u(q)_l} can be cast into the form~\eqref{eq:zrho_l} using the property that
\begin{align}
\delta \rho({\bf q}) & = -\rho\, \delta u^{aa}({\bf q}) =  {\rm i}\, \rho  q \, \delta u_{\rm l}(\bf{q})\, , \label{eq:rho-u0}
\end{align}
holding in the absence of vacancies for the perfect crystal.  The transverse equation~\eqref{eq:u(q)_t} gives equation~\eqref{eq:zu_t}.

\paragraph{From equation~\eqref{eq:fl-T} for the temperature to the longitudinal equation~\eqref{eq:zT_l}.} First, equation~\eqref{eq:rho-u0} can be used to replace $\delta{u}^{aa}$ by $\delta\rho$ into equation~\eqref{eq:fl-T}, giving
\begin{align}
\left( z +\frac{\kappa }{\rho c_v} q^2\right)\delta \tilde T(\mathbf{q},z)- z\frac{\gamma -1}{\rho \alpha}\, \delta \tilde\rho(\mathbf{q},z) & =\delta T(\mathbf{q},0)-\frac{\gamma -1}{\rho \alpha}\, \delta \rho(\mathbf{q},0)   \, .
\end{align}
Next, equation~\eqref{eq:zrho_l} is used to obtain equation~\eqref{eq:zT_l}.

\paragraph{From equation~\eqref{eq:fl-v} for the momentum density to the longitudinal and transverse equations~\eqref{eq:zv_l} and~\eqref{eq:zv_t}.} Contracting equation~\eqref{eq:fl-v} with the vector $e^b_\sigma$ and using the decomposition~\eqref{eq:vlt} leads to
\begin{align}
z \rho \, \delta\tilde v_\sigma(\mathbf{q},z) +\eta^{abcd} q^a q^c e^b_\sigma \, \delta \tilde v^d(\mathbf{q},z) +{\rm i} B^{abcd}_T q^a e^b_\sigma \, \delta \tilde u^{cd}(\mathbf{q},z)- {\rm i}  \alpha B_T \, q \, \delta_{{\rm l}\sigma} \, \delta \tilde T(\mathbf{q},z)  &= \rho\,  \delta v_\sigma(\mathbf{q},0) \, ,
\label{eq:fl-v_sigma1}
\end{align}
since $q^be^b_\sigma = q \, e^b_{\rm l}e^b_\sigma = q \, \delta_{{\rm l}\sigma}$.  With the further decomposition~\eqref{eq:uablt} and the symmetries $ B^{abcd}_T= B^{abdc}_T$ and $ \eta^{abcd}= \eta^{abdc}$, we find that
\begin{align}
{\rm i} B^{abcd}_T q^a e^b_\sigma \, \delta \tilde u^{cd} & = q^2 \sum_{\sigma^\prime} B^T_{\sigma\sigma^\prime} \, \delta\tilde u_{\sigma^\prime} \, , \\
\eta^{abcd} q^a q^c e^b_\sigma \, \delta \tilde v^d & = q^2 \sum_{\sigma^\prime} \eta_{\sigma\sigma^\prime} \, \delta\tilde v_{\sigma^\prime} \, , 
\end{align}
as expressed in terms of the rank-two tensors defined in equation~\eqref{eq:3x3_B+eta}.  Therefore, equation~\eqref{eq:fl-v_sigma1} becomes
\begin{align}
z \rho \, \delta\tilde v_\sigma(\mathbf{q},z) +q^2 \sum_{\sigma^\prime} \eta_{\sigma\sigma^\prime} \, \delta\tilde v_{\sigma^\prime}(\mathbf{q},z) +q^2 \sum_{\sigma^\prime} B^T_{\sigma\sigma^\prime} \, \delta\tilde u_{\sigma^\prime}(\mathbf{q},z)- {\rm i}  \alpha B_T \, q \, \delta_{{\rm l}\sigma} \, \delta \tilde T(\mathbf{q},z)  &= \rho\,  \delta v_\sigma(\mathbf{q},0) \, .
\label{eq:fl-v_sigma2}
\end{align}

As shown here below, the symmetric tensors $B^T_{\sigma\sigma^\prime}$ and $\eta_{\sigma\sigma^\prime}$ can be simultaneously diagonalized if the wave vector is oriented in the special directions of table~\ref{Tab:LTCoeffs}, where $B^T_{\sigma\sigma^\prime}=B^T_{\sigma}\, \delta_{\sigma\sigma^\prime}$ and $\eta_{\sigma\sigma^\prime}=\eta_{\sigma}\, \delta_{\sigma\sigma^\prime}$.  Under such circumstances, the three coupled equations~\eqref{eq:fl-v_sigma2} split into the following three decoupled equations,
\begin{align}
\left( z \rho + \eta_\sigma q^2 \right) \delta\tilde v_{\sigma}(\mathbf{q},z) + B^T_\sigma q^2 \, \delta\tilde u_{\sigma}(\mathbf{q},z)- {\rm i}  \alpha B_T \, q \, \delta_{{\rm l}\sigma} \, \delta \tilde T(\mathbf{q},z)  &= \rho\,  \delta v_\sigma(\mathbf{q},0) \, ,
\label{eq:fl-v_sigma2_diag}
\end{align}
and there is no Einstein’s summation for the indices $\sigma$.  On the one hand, setting $\sigma={\rm l}$, dividing by $\rho$, and using $\delta\tilde u_{\rm l} = -{\rm i} \,\delta\tilde\rho/(\rho  q)$, the longitudinal equation~\eqref{eq:zv_l} is obtained.   On the other hand, setting $\sigma={{\rm t}_k}$ and dividing by $\rho$, the transverse equation~\eqref{eq:zv_t} is found.

\paragraph{Simultaneous diagonalization of the rank-two tensors $B^T_{\sigma\sigma^\prime}$ and $\eta_{\sigma\sigma^\prime}$.} These rank-two tensors form two real symmetric $3\times 3$ matrices ${\mathsf L}=[B^T_{\sigma\sigma^\prime}]$ and ${\mathsf H}=[\eta_{\sigma\sigma^\prime}]$.  The condition to simultaneously diagonalize these two matrices is that they commute.  Since the unit vectors ${\bf e}_\sigma$ form an orthonormal basis, the $3\times 3$ matrix ${\mathsf O}=[ e^a_\sigma]$ with $a=x,y,z$ and $\sigma={\rm l},{\rm t}_1,{\rm t}_2$ defines an orthogonal transformation.  Now, two commuting matrices that undergo the same orthogonal transformation are also commuting.  Therefore, because of the definitions~\eqref{eq:3x3_B+eta} of these matrices, we may equivalently consider the following $3\times 3$ matrices, 
\begin{align}
\left[ B^{abcd}_T q^a q^c \right] &= 
\left[
\begin{array}{ccc}
B^T_{11} q_x^{2} + B^T_{44}(q_y^{2}+q_z^{2}) & (B^T_{12}+B^T_{44}) q_{x} q_{y} & (B^T_{12}+B^T_{44}) q_{x} q_{z} \\
(B^T_{12}+B^T_{44}) q_{x} q_{y} & B^T_{11} q_y^{2} + B^T_{44}(q_x^{2}+q_z^{2}) & (B^T_{12}+B^T_{44}) q_{y} q_{z} \\
(B^T_{12}+B^T_{44}) q_{x} q_{z} & (B^T_{12}+B^T_{44}) q_{y} q_{z} & B^T_{11} q_z^{2} + B^T_{44}(q_x^{2}+q_y^{2}) \\
\end{array}
\right] , \label{eq:matrix-B^T}\\
\left[ \eta^{abcd} q^a q^c \right] & = 
\left[
\begin{array}{ccc}
\eta_{11} q_x^{2} + \eta_{44}(q_y^{2}+q_z^{2}) & (\eta_{12}+\eta_{44}) q_{x} q_{y} & (\eta_{12}+\eta_{44}) q_{x} q_{z} \\
(\eta_{12}+\eta_{44}) q_{x} q_{y} & \eta_{11} q_y^{2} + \eta_{44}(q_x^{2}+q_z^{2}) & (\eta_{12}+\eta_{44}) q_{y} q_{z} \\
(\eta_{12}+\eta_{44}) q_{x} q_{z} & (\eta_{12}+\eta_{44}) q_{y} q_{z} & \eta_{11} q_z^{2} + \eta_{44}(q_x^{2}+q_y^{2}) \\
\end{array}
\right] , \label{eq:matrix-eta}
\end{align}
which are given for a cubic crystal using Voigt's notations.  The conditions for these two matrices to commute are that
\begin{align}
q_x \, q_y \left(q_x^2-q_y^2\right) = 0 \, , \qquad
q_y \, q_z \left(q_y^2-q_z^2\right) = 0 \, , \qquad\mbox{and}\qquad
q_z \, q_x \left(q_z^2-q_x^2\right) = 0 \, .
\end{align}
These conditions imply that the wave vector ${\bf q}$ should be oriented in one of the directions $[100]$, $[110]$, $[111]$, or the symmetry-related directions obtained by the reflections of $1$ into $\bar{1}$ and by the cyclic permutations of the three indices.  They are the special directions considered in table~\ref{Tab:LTCoeffs}.

In the direction $[100]$, the wave vector is given by ${\bf q}=(q,0,0)$ and the matrices~\eqref{eq:matrix-B^T} and~\eqref{eq:matrix-eta} have the following forms,
\begin{align}
\left[ B^{abcd}_T q^a q^c \right] &=  q^2 
\left[
\begin{array}{ccc}
B^T_{11} & 0 & 0 \\
0& B^T_{44} & 0 \\
0 & 0 & B^T_{44} \\
\end{array}
\right] ,
\qquad
\left[ \eta^{abcd} q^a q^c \right] =  q^2 
\left[
\begin{array}{ccc}
\eta_{11} & 0 & 0 \\
0& \eta_{44} & 0 \\
0 & 0 & \eta_{44} \\
\end{array}
\right] , \label{eq:matrices-100}
\end{align}
which are directly diagonal, leading to the corresponding eigenvalues and eigenvectors given in table~\ref{Tab:LTCoeffs}.

In the direction $[110]$, the wave vector is given by ${\bf q}=(q,q,0)/\sqrt{2}$ and the matrices~\eqref{eq:matrix-B^T} and~\eqref{eq:matrix-eta} have the following forms,
\begin{align}
\left[ B^{abcd}_T q^a q^c \right] &=  \frac{q^2}{2}
\left[
\begin{array}{ccc}
B^T_{11}+B^T_{44} & B^T_{12}+B^T_{44} & 0 \\
B^T_{12}+B^T_{44} & B^T_{11}+B^T_{44} & 0 \\
0 & 0 & 2 \, B^T_{44} \\
\end{array}
\right] ,
\qquad
\left[ \eta^{abcd} q^a q^c \right] =  \frac{q^2}{2}
\left[
\begin{array}{ccc}
\eta_{11}+\eta_{44} & \eta_{12}+\eta_{44} & 0 \\
\eta_{12}+\eta_{44} & \eta_{11}+\eta_{44} & 0 \\
0 & 0 & 2 \, \eta_{44} \\
\end{array}
\right] , \label{eq:matrices-110}
\end{align}
which commute and can thus be simultaneously diagonalized, leading to the corresponding eigenvalues and eigenvectors given in table~\ref{Tab:LTCoeffs}.

In the direction $[111]$, the wave vector is given by ${\bf q}=(q,q,q)/\sqrt{3}$ and the matrices~\eqref{eq:matrix-B^T} and~\eqref{eq:matrix-eta} have the following forms,
\begin{align}
\left[ B^{abcd}_T q^a q^c \right] &=  \frac{q^2}{3}
\left[
\begin{array}{ccc}
B^T_{11}+2\, B^T_{44} & B^T_{12}+B^T_{44} & B^T_{12}+B^T_{44} \\
B^T_{12}+B^T_{44} & B^T_{11}+2\, B^T_{44} & B^T_{12}+B^T_{44} \\
B^T_{12}+B^T_{44} & B^T_{12}+B^T_{44} & B^T_{11}+2\, B^T_{44} \\
\end{array}
\right] ,
\notag\\
\left[ \eta^{abcd} q^a q^c \right] &=  \frac{q^2}{3}
\left[
\begin{array}{ccc}
\eta_{11}+2\, \eta_{44} & \eta_{12}+\eta_{44} & \eta_{12}+\eta_{44} \\
\eta_{12}+\eta_{44} & \eta_{11}+2\, \eta_{44} & \eta_{12}+\eta_{44} \\
\eta_{12}+\eta_{44} & \eta_{12}+\eta_{44} & \eta_{11}+2 \, \eta_{44} \\
\end{array}
\right] , \label{eq:matrices-110}
\end{align}
which commute and can thus be simultaneously diagonalized, leading to the corresponding eigenvalues and eigenvectors given in table~\ref{Tab:LTCoeffs}.

Therefore, the three coupled equations~\eqref{eq:fl-v_sigma2} split into the three decoupled equations~\eqref{eq:fl-v_sigma2_diag}.

%%%%%%%%%%%%%%%%%%%%%%%%%%%%%%%%%%%%%%%%%%%%%%%%%%%%%%%%%%%%
\subsection{Calculations of the longitudinal and transverse correlation and spectral functions}
\label{app:calc_tlc}
%%%%%%%%%%%%%%%%%%%%%%%%%%%%%%%%%%%%%%%%%%%%%%%%%%%%%%%%%%%%
\paragraph{Full expression of the dynamic structure factor equation~\eqref{eq:DSF_Full}.} Using equation~\eqref{eq:FTtoLT}, the dynamic structure factor is obtained  from its Laplace transform as $S({\bf q},\omega)  = 2\, S({\bf q})\,{\rm Re}\,  \tilde F({\bf{q}},z={\rm i} \omega)$. Setting  $z={\rm i} \omega$ in the numerator of right-hand side of equation~\eqref{eq:Sks} gives
\begin{align}
( {\rm i}\omega+D_vq^2)( {\rm i}\omega+\gamma D_T q^2)+\frac{\left(\gamma-1\right)B_T}{\rho}q^2 &=-\omega^2+\frac{\left(\gamma-1\right)B_T}{\rho}q^2+\gamma D_T D_v q^4+{\rm i}\omega (D_v+\gamma D_T)q^2\notag\\
&=N_1(\omega)+{\rm i} N_2(\omega)\, ,
\end{align}
which defines~\eqref{N1} and~\eqref{N2}.
Moreover, the determinant of the matrix $\boldsymbol{\mathsf M}$ is
\begin{align}
\det \boldsymbol{\mathsf M}({q},z)&=z^3+z^2(D_v+\gamma D_T)q^2+z\left[\frac{B^T_{\rm l}}{\rho} q^2+\frac{\left(\gamma-1\right)B_T}{\rho}q^2+\gamma D_T D_v q^4\right]+\frac{\gamma B^T_{\rm l}D_T}{\rho}q^4 \, ,\label{eq:detM}
\end{align}
and we obtain
\begin{align}
	\det \boldsymbol{\mathsf M}({q},z={\rm i} \omega)&=-{\rm i}\omega^3-\omega^2(D_v+\gamma D_T)q^2+{\rm i}\omega\left[\frac{B^T_{\rm l}}{\rho}q^2+\frac{\left(\gamma-1\right)B_T}{\rho}q^2+\gamma D_T D_v q^4\right]+\frac{\gamma B^T_{\rm l}D_T}{\rho}q^4\notag\\
	&=-\omega^2(D_v+\gamma D_T)q^2+\frac{\gamma B^T_{\rm l}D_T}{\rho}q^4+{\rm i}\omega\left[-\omega^2+\frac{B^T_{\rm l}}{\rho}q^2+\frac{\left(\gamma-1\right)B_T}{\rho}q^2+\gamma D_T D_v q^4\right]\notag\\
	&= D_1(\omega)+ {\rm i} D_2(\omega)\, ,
\end{align}
which defines~\eqref{D1} and~\eqref{D2}.
Equation~\eqref{eq:DSF_Full} is obtained from
\begin{align}
	\frac{S({q},\omega)}{S({q})}&=2\, {\rm Re}\, \frac{N_1(\omega)+{\rm i} N_2(\omega)}{D_1(\omega)+ {\rm i} D_2(\omega)}\frac{D_1(\omega)- {\rm i} D_2(\omega)}{D_1(\omega)-{\rm i} D_2(\omega)} =2 \, \frac{N_1(\omega)D_1(\omega)+N_2(\omega)D_2(\omega)}{D^2_1(\omega)+D^2_2(\omega)}\, .
\end{align}

\paragraph{Rayleigh and Brillouin peaks of the dynamic structure factor~\eqref{eq:DSF_RBP}.}
The intermediate scattering function $F({q},t)$ is obtained from the inverse Laplace transform of equation~\eqref{eq:Sks}.  Its denominator is a cubic polynomial, which can be factorized as $\det \boldsymbol{\mathsf M}({q},z)=\prod_{j=0,{\rm l}\pm}(z-z_j)$ in terms of the roots given by
\begin{align}
z_0 &= -\chi q^2+\cdots\, , && z_{{\rm l}\pm} = \pm {\rm i} c_{\rm l} q - \Gamma_{\rm l} q^2+\cdots\, .\label{eq:polesl}
\end{align} 
Therefore,  the intermediate scattering function can be calculated for $t>0$ as
\begin{align}
\frac{F({q},t)}{S({q})}&=\frac{1}{2\pi {\rm i}}\int_{c-{\rm i}\infty}^{c+{\rm i}\infty}\text{d}z \, {\rm e}^{zt} \, \frac{G(z)}{\prod_{j=0,{\rm l}\pm}(z-z_j)}=\sum_{k=0,{\rm l}\pm} {\rm e}^{z_kt} \lim_{z\rightarrow z_k}\frac{G(z)\, (z-z_k)}{\prod_{j=0,{\rm l}\pm}(z-z_j)} \, ,\label{eq:DSFstILT}
\end{align}
where
\begin{align}
G(z) & \equiv ( z+D_vq^2)(z+\gamma D_T q^2)+\frac{\left(\gamma-1\right) B_T}{\rho}q^2\, .
\end{align}

The computation of the $j=0$ term in the sum in equation~\eqref{eq:DSFstILT} gives
\begin{align}
G(z_{0}) & =\left( D_v  -\chi \right)\left(\gamma D_T -\chi \right)  q^4 +\frac{(\gamma-1)B_T}{\rho}q^2=\frac{(\gamma-1)B_T}{\rho}q^2+\mathcal{O}(q^4)\, ,\\
\lim_{z\rightarrow z_{0}}\frac{(z-z_0)}{\det \boldsymbol{\mathsf M}({q},z)}&=
\left[\left(-\chi q^2- {\rm i} c_{\rm l} q + \Gamma_{\rm l} q^2\right)\left(-\chi q^2 + {\rm i} c_{\rm l} q + \Gamma_{\rm l} q^2\right)\right]^{-1}=\left(c_{\rm l} q\right)^{-2}+\mathcal{O}(q^0)\, ,
\end{align}
and, at leading order in $q$, we thus find
\begin{align}
{\rm e}^{z_0t} \lim_{z\rightarrow z_{0}}\frac{G(z)\,(z-z_{0})}{\det \boldsymbol{\mathsf M}({q},z)} &=\frac{\gamma-1}{\rho}\frac{B_T}{c^2_{\rm l}}\, {\rm e}^{-\chi q^2t} =\frac{1}{1+\frac{1}{\gamma-1}\frac{B^T_{\rm l}}{B_T}} \, {\rm e}^{-\chi q^2t} \, .\label{eq:iltz0}
\end{align}

For the $j={\rm l}+$ term in the sum in equation~\eqref{eq:DSFstILT}, we have
\begin{align}
G(z_{{\rm l}+})&=\left[{\rm i} c_{\rm l}q+(D_v-\Gamma_{\rm l}) q^2\right]\left[ {\rm i} c_{\rm l}q+(\gamma D_T-\Gamma_{\rm l}) q^2\right]+\frac{(\gamma-1)B_T}{\rho}q^2\notag\\
	&= \left[-c_{\rm l}^2+\frac{(\gamma-1)B_T}{\rho}\right]q^2+{\rm i} c_{\rm l}q^3(D_v+\gamma D_T-2\Gamma_{\rm l})+ q^4(D_v-\Gamma_{\rm l})(\gamma D_T-\Gamma_{\rm l})\notag\\
	&=-\frac{B^T_{\rm l}}{\rho}q^2+{\rm i} c_{\rm l}q^3(D_v+\gamma D_T-2\Gamma_{\rm l})+  \mathcal{O}(q^4)\notag\\
	&=-\frac{B^T_{\rm l}}{\rho}q^2+{\rm i} \frac{\gamma B^T_{\rm l} D_T}{\rho c_{\rm l}}q^3+  \mathcal{O}(q^4)\, ,\\
	\lim_{z\rightarrow z_{{\rm l}+}}\frac{(z-z_{{\rm l}+})}{\det \boldsymbol{\mathsf M}({q},z)}&=\left[\left( {\rm i} c_{\rm l}q-\Gamma_{\rm l} q^2 + \chi q^2\right)\left({\rm i} c_{\rm l}q-\Gamma_{\rm l} q^2+{\rm i}c_{\rm l}q+\Gamma_{\rm l} q^2\right)\right]^{-1}\notag\\
	&=\left[-2c_{\rm l}^2q^2+2{\rm i}c_{\rm l}\left(\chi -\Gamma_{\rm l}\right)q^3 +\mathcal{O}(q^4)\right]^{-1}\, .
\end{align}
At leading orders in $q$, we obtain
\begin{align}
{\rm e}^{z_{{\rm l}+}t}\lim_{z\rightarrow z_{{\rm l}+}}\frac{G(z)\,(z-z_{{\rm l}+})}{\det \boldsymbol{\mathsf M}({q},z)}&=\frac{-\frac{B^T_{\rm l}}{\rho}q^2+{\rm i} \frac{\gamma B^T_{\rm l} D_T}{\rho c_{\rm l}}q^3+  \mathcal{O}(q^4)}{-2c_{\rm l}^2q^2\left[1-\frac{{\rm i} q}{c_{\rm l}}\left(\chi-\Gamma_{\rm l}\right) +\mathcal{O}(q^2)\right]}\, {\rm e}^{\left({\rm i} c_{\rm l}q-\Gamma_{\rm l} q^2\right)t}\notag\\
&=\frac{B^T_{\rm l}}{2\rho c^2_{\rm l}}\left[1- \frac{{\rm i} q}{c_{\rm l}}\gamma D_T+  \mathcal{O}(q^2)\right]\left[1+\frac{{\rm i} q}{c_{\rm l}}\left(\chi-\Gamma_{\rm l}\right) +\mathcal{O}(q^2)\right]{\rm e}^{\left({\rm i} c_{\rm l}q-\Gamma_{\rm l} q^2\right)t}\notag\\
&=\frac{B^T_{\rm l}}{2\rho c^2_{\rm l}}\left[1+\frac{{\rm i} q}{c_{\rm l}}\left(\chi -\Gamma_{\rm l}-\gamma D_T\right) +\mathcal{O}(q^2)\right]{\rm e}^{\left({\rm i} c_{\rm l}q-\Gamma_{\rm l} q^2\right)t}\notag\\
&=\frac{1}{2\left[1+(\gamma-1)\frac{B_T}{B^T_{\rm l}}\right]}\left[1+\frac{{\rm i} q}{c_{\rm l}}\left(D_v-3\Gamma_{\rm l}\right) +\mathcal{O}(q^2)\right]{\rm e}^{\left({\rm i} c_{\rm l}q-\Gamma_{\rm l} q^2\right)t}\,,\label{eq:iltzp}
\end{align}
using the fact that
\begin{align}
\chi-\gamma D_T & = D_v-2\Gamma_{\rm l}\, .
\end{align}

Similarly, for $j={\rm l}-$, we obtain
\begin{align}
{\rm e}^{z_{{\rm l}-}t}\lim_{z\rightarrow z_{{\rm l}-}}\frac{G(z)\,(z-z_{{\rm l}-})}{\det \boldsymbol{\mathsf M}({q},z)}&=\frac{1}{2\left[1+(\gamma-1)\frac{B_T}{B^T_{\rm l}}\right]}\left[1-\frac{{\rm i} q}{c_{\rm l}}\left(D_v-3\Gamma_{\rm l}\right) +\mathcal{O}(q^2)\right]{\rm e}^{\left(-{\rm i} c_{\rm l}q-\Gamma_{\rm l} q^2\right)t}\,.\label{eq:iltzm}
\end{align}
Substituting equations~\eqref{eq:iltz0},~\eqref{eq:iltzp}, and~\eqref{eq:iltzm} into equation~\eqref{eq:DSFstILT}  and taking into account the property~\eqref{eq:C:microrev} gives the intermediate scattering function~\eqref{eq:ISF}.

\paragraph{Solutions of equations~\eqref{eq:zv_t} and~\eqref{eq:zu_t}.} The set of these transverse equations are cast in a matrix form  as
\begin{align}\label{eq:lin-vti-ui}
\left[
\begin{array}{ll}
z+ \frac{\eta_{{\rm t}_k}}{\rho}q^2 &  \frac{B^T_{{\rm t}_k}}{\rho}q^2    \\
-1 & z
\end{array}
\right]
\left[
\begin{array}{l}
\delta \tilde v_{{\rm t}_k}({\bf q},z) \\
\delta \tilde u_{{\rm t}_k}({\bf q},z)\\
\end{array}
\right]
=
\left[
\begin{array}{l}
\delta v_{{\rm t}_k}({\bf q},0) \\
\delta u_{{\rm t}_k}({\bf q},0)\\
\end{array}
\right] .
\end{align}
The solution is obtained by matrix inversion
\begin{align}
\left[
\begin{array}{l}
\delta \tilde  v_{{\rm t}_k}({\bf q},z) \\
\delta\tilde  u_{{\rm t}_k}({\bf q},z)\\
\end{array}
\right]
&=
\frac{1}{z\left(z+ \frac{\eta_{{\rm t}_k}}{\rho}q^2\right)+ \frac{B^T_{{\rm t}_k}}{\rho}q^2}
\left[
\begin{array}{ll}
z &  -\frac{B^T_{{\rm t}_k}}{\rho}q^2    \\
1 & z+ \frac{\eta_{{\rm t}_k}}{\rho}q^2
\end{array}
\right]
\left[
\begin{array}{l}
\delta v_{{\rm t}_k}({\bf q},0) \\
\delta u_{{\rm t}_k}({\bf q},0)\\
\end{array}
\right]\notag\\
&=\frac{1}{z\left(z+ \frac{\eta_{{\rm t}_k}}{\rho}q^2\right)+\frac{B^T_{{\rm t}_k}}{\rho}q^2}
\left[
\begin{array}{l}
z\, \delta v_{{\rm t}_k}({\bf q},0)-\frac{B^T_{{\rm t}_k}}{\rho}q^2\, \delta u_{{\rm t}_k}({\bf q},0) \\
\delta v_{{\rm t}_k}({\bf q},0)+\left(z+ \frac{\eta_{{\rm t}_k}}{\rho}q^2\right)\delta u_{{\rm t}_k}({\bf q},0)\\
\end{array}
\right] .
\label{eq:vutz}
\end{align}
Using the hypothesis of the regression of fluctuations~\cite{MG23},  the deviations $\delta v_{{\rm t}_k}$ and $\delta u_{{\rm t}_k}$ are replaced by their microscopic expressions $\delta \hat{v}_{{\rm t}_k}$ and $\delta \hat{u}_{{\rm t}_k}$. The Laplace transforms of the correlation functions are thus obtained by multiplying $\delta \hat{\tilde v}_{{\rm t}_k}$ and $\delta \hat{\tilde u}_{{\rm t}_k}$ given by equation~\eqref{eq:vutz} by  $\delta \hat{v}^*_{{\rm t}_k}$ and $\delta \hat{u}^*_{{\rm t}_k}$ on the right-hand side and taking the statistical average $\langle\cdot\rangle_{\rm eq}$ with respect to the equilibrium probability distribution.  Since $\langle\delta \hat{v}_{{\rm t}_k}({\bf q},0)\,\delta \hat{u}^*_{{\rm t}_k}({\bf q},0)\rangle_{\rm eq}=0$, the equations~\eqref{eq:vvwovvt} and~\eqref{eq:uuwouut} are obtained.

%%%%%%%%%%%%%%%%%%%%%%%%%%%%%%%%%%%%%%%%%%%%%%%%%%%%%%%%%%%%%%%%%%

%%%%%%%%%%%%%%%%%%%%%%%%%%%%%%%%%%%%%%%%%%%%%%%%%%%%%%%%%%%%%%%%%%%%%%%
\pagebreak
%%%%%%%%%%%%%%%%%%%%%%%%%%%%%%%%%%%%%%%%%%%%%%%%%%%%%%%%%%%%%%%%%%%%%%%

\begin{table}[h!]
  \begin{tabular}{c @{\hskip .5cm}c @{\hskip .5cm}c@{\hskip .5cm}c@{\hskip .5cm}c @{\hskip .5cm}c}
    \hline\hline
    $n_*$ & $\chi_*$ &$c_{\rm l*}$ & $\Gamma_{\rm l*}$ & $ c_{{\rm t}_{1,2}*}$ & $ \Gamma_{{\rm t}_{1,2}*}$     \\ 
    \hline
    1.037 & $3.56\pm0.16$	&$11.86\pm0.04$	&$4.10\pm0.12$	&$5.64\pm0.18$	&$2.03\pm0.03$ \\
    1.1 &$4.58\pm0.21$  &$14.51\pm0.02$	&$4.55\pm0.14$	&$7.15\pm0.11$	&$2.44\pm0.07$    \\
    1.2 &$6.97\pm0.38$	&$21.82\pm0.02$	&$6.24\pm0.22$&	$11.45\pm0.15$& $3.61\pm0.12$    \\
    1.3 &$13.49\pm0.39$ &$41.97\pm0.06$&$11.24\pm0.22$&$23.09\pm0.12$&$6.91\pm0.20$ \\
    1.4 &$111.41\pm2.43$ &$345.07\pm0.44$&$89.32\pm1.75$&$198.84\pm1.89$&$55.34\pm1.86$ 
     \\      \hline\hline
  \end{tabular}
  \caption{The coefficients of the dispersion relations~\eqref{eq:DRl} and~\eqref{eq:poles_Jtqw} versus the density $n_*$ for the wave vector ${\bf q}$ in the direction~$[100]$:  The coefficient $\chi$ is given by equation~\eqref{eq:thdi}, the speed of the longitudinal sound wave $c_{\rm l }$ by equation~\eqref{eq:cl}, the longitudinal acoustic attenuation coefficient $\Gamma_{\rm l}$  by equation~\eqref{eq:Gaml}, and the speeds of the transverse sound waves $c_{{\rm t}_k}$ and the transverse acoustic attenuation coefficients $\Gamma_{{\rm t}_k}$ by equation~\eqref{eq:c_t+G_t}. The data for the thermodynamic, elastic, and transport properties are taken from reference~\cite{MG23_primo}.}\label{Tab:PC100}
\end{table}

%%%%%%%%%%%%%%%%%%%%%%%%%%%%%%%%%%%%%%%%%%%%%%%%%%%%%%%%%%%%%%%%%%%%%%%

\begin{table}[h!]
  \begin{tabular}{c @{\hskip .5cm}c @{\hskip .5cm}c@{\hskip .5cm}c@{\hskip .5cm}c @{\hskip .5cm}c@{\hskip .5cm}c@{\hskip .5cm}c}
    \hline\hline
    $n_*$ & $\chi_*$ &$c_{\rm l*}$ & $\Gamma_{\rm l*}$ & $ c_{{\rm t}_1*}$ & $ \Gamma_{{\rm t}_1*}$ &  $ c_{{\rm t}_2*}$ & $ \Gamma_{{\rm t}_2*}$    \\ 
    \hline
    1.037 & $4.13\pm0.19$ & $12.59\pm0.08$ & $4.80\pm0.12$ & $3.73\pm0.06$ & $1.04\pm0.04$ & $5.64\pm0.18$ & $2.03\pm0.03$ \\
    1.1 &$5.20\pm0.24$ &$15.36\pm0.05$	&$5.48\pm0.14$	&$5.06\pm0.03$	&$1.20\pm0.04$& $7.15\pm0.11$& $2.44\pm0.07$    \\
    1.2 &$7.85\pm0.43$&$23.17\pm0.08$&	$7.69\pm0.23$&$8.40\pm0.02$&$1.72\pm0.04$&$11.45\pm0.15$&$3.61\pm0.12$    \\
    1.3 &$15.13\pm0.44$ &$44.65\pm0.07$&$14.19\pm0.28$&$17.37\pm0.08$&$3.14\pm0.05$&$23.09\pm0.12$&$6.91\pm0.20$ \\
    1.4  &$125.12\pm2.77$ &$368.77\pm1.04$&$112.15\pm2.30$&$150.39\pm0.57$&$25.66\pm0.87$&$198.84\pm1.89$&$55.34\pm1.86$ 
     \\      \hline\hline
  \end{tabular}
  \caption{The coefficients of the dispersion relations~\eqref{eq:DRl} and~\eqref{eq:poles_Jtqw} versus the density $n_*$ for the wave vector ${\bf q}$ in the direction~$[110]$:  The coefficient $\chi$ is given by equation~\eqref{eq:thdi}, the speed of the longitudinal sound wave $c_{\rm l }$ by equation~\eqref{eq:cl}, the  longitudinal acoustic attenuation coefficient $\Gamma_{\rm l}$  by equation~\eqref{eq:Gaml}, and the speeds of the transverse sound waves $c_{{\rm t}_k}$  and the transverse acoustic attenuation coefficients $\Gamma_{{\rm t}_k}$ by equation~\eqref{eq:c_t+G_t}. The data for the thermodynamical, elastic, and transport coefficients are taken from reference~\cite{MG23_primo}.}\label{Tab:PC110}
\end{table}

%%%%%%%%%%%%%%%%%%%%%%%%%%%%%%%%%%%%%%%%%%%%%%%%%%%%%%%%%%%%%%%%%%%%%%%

\begin{table}[h!]
  \begin{tabular}{c @{\hskip .5cm}c @{\hskip .5cm}c@{\hskip .5cm}c@{\hskip .5cm}c @{\hskip .5cm}c}
    \hline\hline
    $n_*$ & $\chi_*$ &$c_{\rm l*}$ & $\Gamma_{\rm l*}$ & $ c_{{\rm t}_{1,2}*}$ & $ \Gamma_{{\rm t}_{1,2}*}$    \\ 
    \hline
    1.037 & $4.30\pm0.21$& $12.83\pm0.11$ & $5.04\pm0.11$&$4.46\pm0.09$	&$1.37\pm0.03$ \\
    1.1 &$5.38\pm0.25$ &	$15.63\pm0.07$&	$5.80\pm0.15$& $5.84\pm0.05$& $1.61\pm0.04$    \\
    1.2 &$8.11\pm0.44$	&$23.60\pm0.10$	&$8.20\pm0.24$	&$9.53\pm0.06$	&$2.35\pm0.05$    \\
    1.3 &$15.59\pm0.45$ &$45.50\pm0.09$&$15.21\pm0.33$&$19.46\pm0.07$&$4.40\pm0.08$ \\
    1.4  & $128.96\pm2.87$&$376.34\pm1.34$&$120.12\pm2.82$&$168.10\pm0.82$&$35.55\pm0.85$ 
     \\      \hline\hline
  \end{tabular}
  \caption{The coefficients of the dispersion relations~\eqref{eq:DRl} and~\eqref{eq:poles_Jtqw} versus the density $n_*$ for the wave vector ${\bf q}$ in the direction~$[111]$:  The coefficient $\chi$ is given by equation~\eqref{eq:thdi}, the speed of the longitudinal sound wave $c_{\rm l }$ by equation~\eqref{eq:cl}, the  longitudinal acoustic attenuation coefficient $\Gamma_{\rm l}$  by equation~\eqref{eq:Gaml}, and the speeds of the transverse sound waves $c_{{\rm t}_k}$ and the transverse acoustic attenuation coefficients $\Gamma_{{\rm t}_k}$ by equation~\eqref{eq:c_t+G_t}. The data for the thermodynamical, elastic, and transport coefficients are taken from reference~\cite{MG23_primo}.}\label{Tab:PC111}
\end{table}

%%%%%%%%%%%%%%%%%%%%%%%%%%%%%%%%%%%%%%%%%%%%%%%%%%%%%%%%%%%%%%%%%%%%%%%

%%%%%%%%%%%%%%%%%%%%%%%%%%%%%%%%%%%%%%%%%%%%%%%%%
\begin{figure}[h!]\centering
{\includegraphics[width=1.0\textwidth]{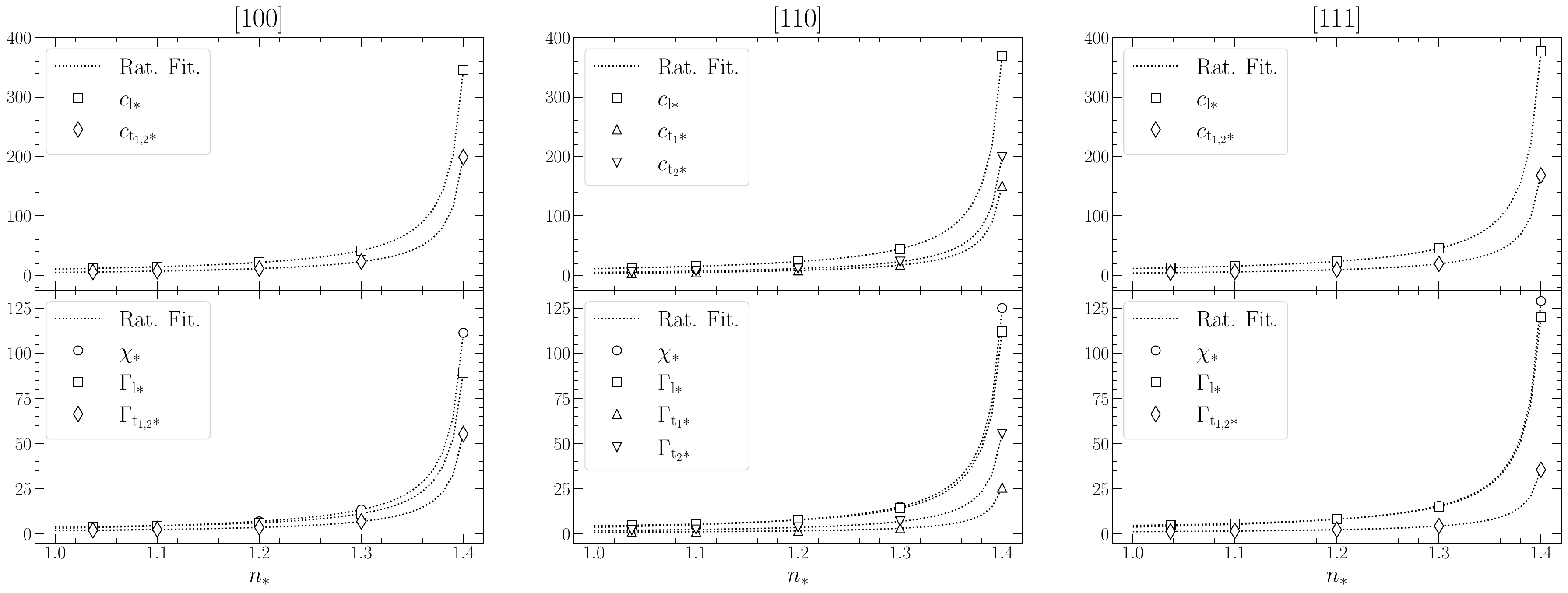}}
\caption[] {The coefficients of tables~\ref{Tab:PC100},~\ref{Tab:PC110}, and ~\ref{Tab:PC111} versus the density $n_*$ for the wave vector ${\bf q}$ in the directions $[100]$, $[110]$, and $[111]$. The coefficient $\chi$ is given by equation~\eqref{eq:thdi}, the speed of the longitudinal sound wave $c_{\rm l }$ by equation~\eqref{eq:cl}, the longitudinal acoustic attenuation coefficient $\Gamma_{\rm l}$  by equation~\eqref{eq:Gaml}, and the speeds of the  transverse sound waves $c_{{\rm t}_k}$ and the transverse acoustic attenuation coefficients $\Gamma_{{\rm t}_k}$ by equation~\eqref{eq:c_t+G_t}.  As a guide to the eye, rational functions have been fitted to the data.}\label{Fig:PC}
\end{figure}

%%%%%%%%%%%%%%%%%%%%%%%%%%%%%%%%%%%%%%%%%%%%%%%%%
\begin{figure}[h!]\centering
{\includegraphics[width=0.75\textwidth]{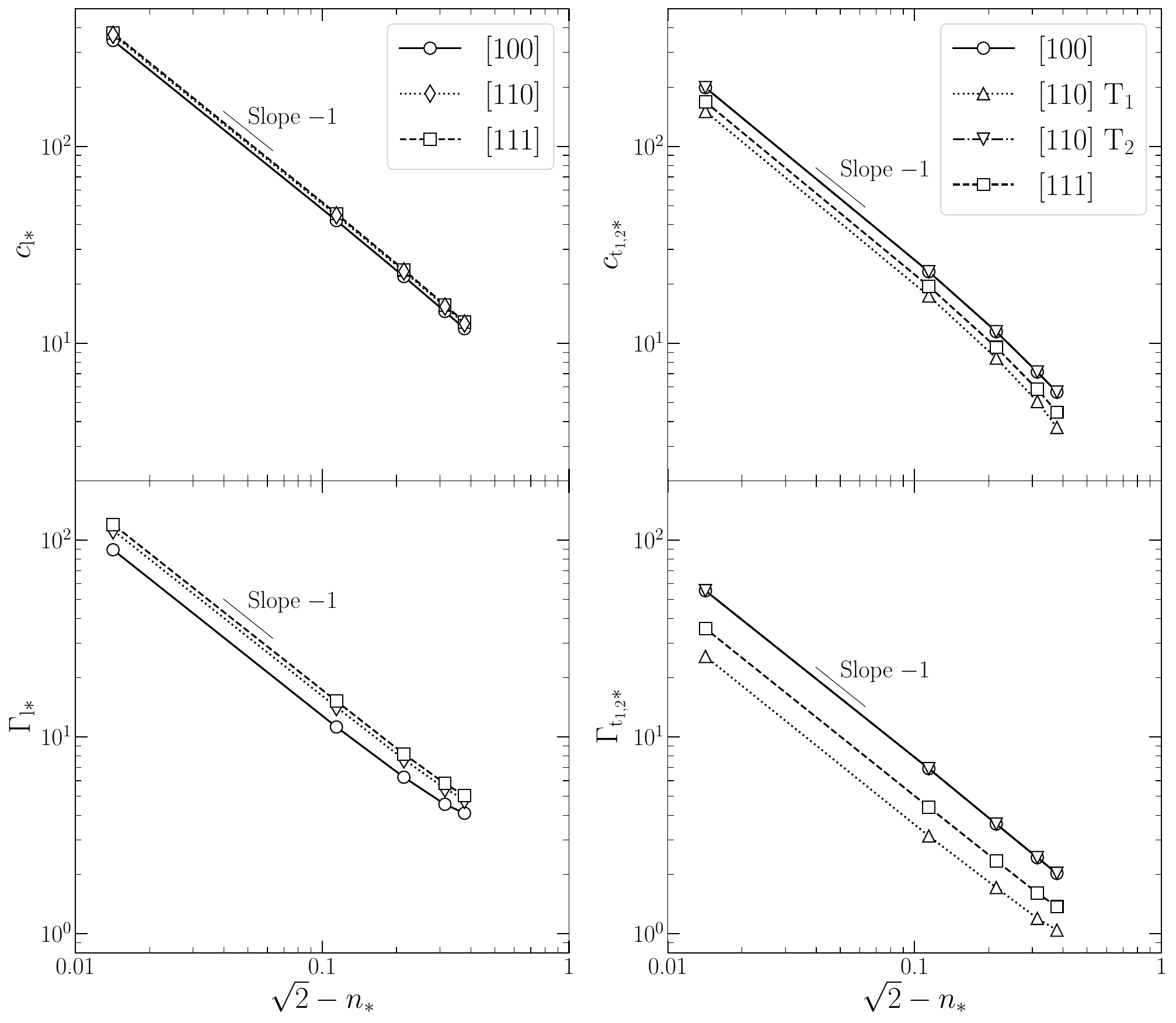}}
\caption[] {Divergences of the speeds and attenuation coefficients of the sound waves  in the vicinity of the close-packing density $n_*=\sqrt{2}$.  The values are given in tables~\ref{Tab:PC100},~\ref{Tab:PC110}, and ~\ref{Tab:PC111} for the speeds of the longitudinal and transverse sound waves $c_{{\rm l}}$~\eqref{eq:cl} and $c_{{\rm t}_k}$~\eqref{eq:c_t+G_t}, and their acoustic coefficients $\Gamma_{\rm l}$~\eqref{eq:Gaml} and $\Gamma_{{\rm t}_k}$~\eqref{eq:c_t+G_t}, versus the density $n_*$ for the wave vector ${\bf q}$ in the directions $[100]$, $[110]$, and $[111]$.}\label{Fig:cs}
\end{figure}

%%%%%%%%%%%%%%%%%%%%%%%%%%%%%%%%%%%%%%%%%%%%%%%%%
\begin{figure}[h!]\centering
{\includegraphics[width=1.0\textwidth]{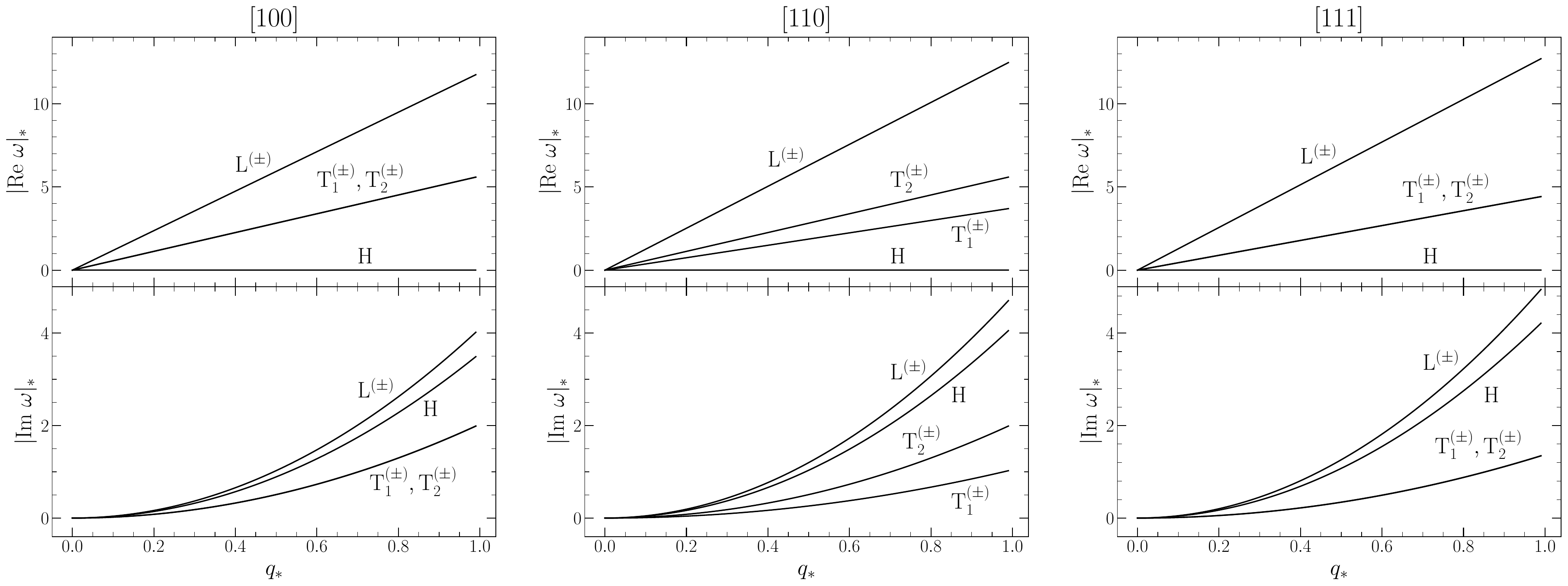}}
\caption[] {The real and imaginary parts of the dispersion relations~\eqref{eq:DRl} and~\eqref{eq:poles_Jtqw} versus the wave number $q_*$ for the hard-sphere crystal at the density $n_*=1.037$ for the wave vector ${\bf q}$ in the directions $[100]$,  $[110]$, and  $[111]$. $\text{L}^{(\pm)}$ denotes the pair of longitudinal sound modes, $\text{T}_{1,2}^{(\pm)}$ the two pairs of transverse sound modes, and $\text{H}$ the heat diffusion mode. The coefficients $\chi$, $c_{\rm l}$, $\Gamma_{\rm l}$, $c_{{\rm t}_k}$, and $\Gamma_{{\rm t}_k}$ used for these plots are taken from tables~\ref{Tab:PC100}-\ref{Tab:PC111}.  }\label{Fig:DR_1d037}
\end{figure}

%%%%%%%%%%%%%%%%%%%%%%%%%%%%%%%%%%%%%%%%%%%%%%%%%

\begin{figure}[h!]\centering
{\includegraphics[width=1.0\textwidth]{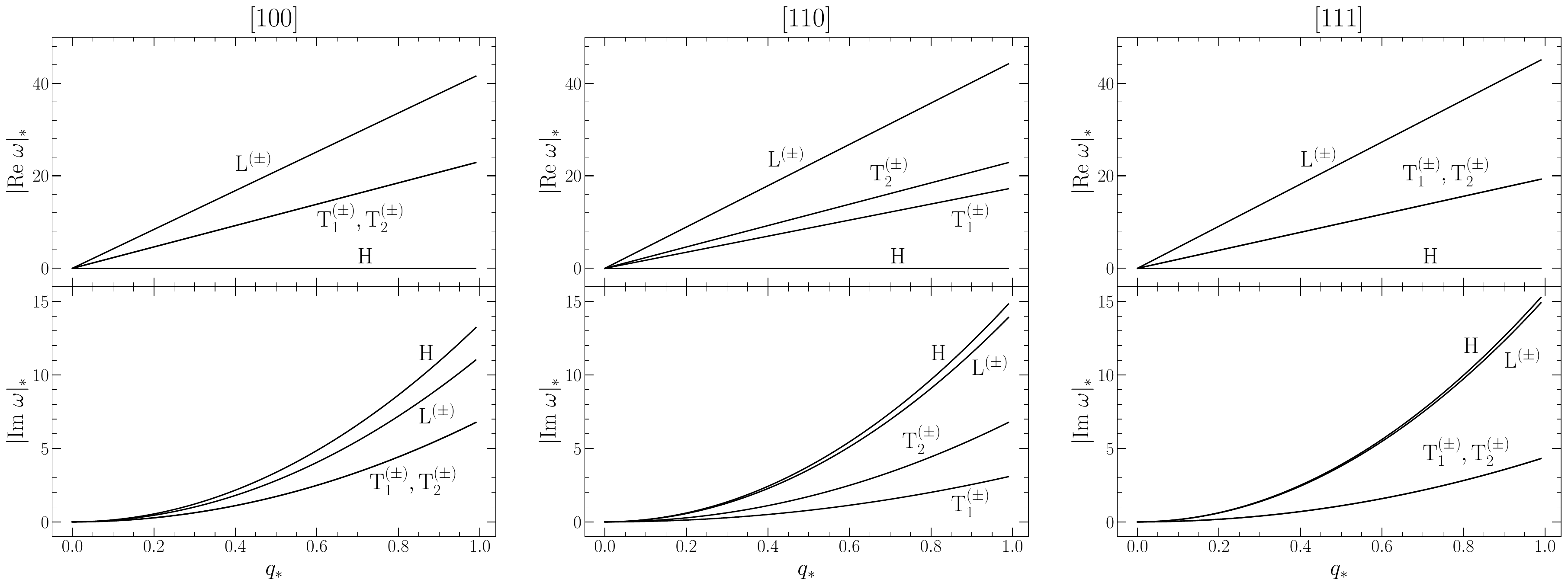}}
\caption[] {The real and imaginary parts of the dispersion relations~\eqref{eq:DRl} and~\eqref{eq:poles_Jtqw} versus the wave number $q_*$ for the hard-sphere crystal at the density $n_*=1.3$ for the wave vector ${\bf q}$ in the directions $[100]$,  $[110]$, and  $[111]$. $\text{L}^{(\pm)}$ denotes the pair of longitudinal sound modes, $\text{T}_{1,2}^{(\pm)}$ the two pairs of transverse sound modes, and $\text{H}$ the heat mode. The coefficients $\chi$, $c_{\rm l}$, $\Gamma_{\rm l}$, $c_{{\rm t}_k}$, and $\Gamma_{{\rm t}_k}$ used for these plots are taken from tables~\ref{Tab:PC100}-\ref{Tab:PC111}.  }\label{Fig:DR_1d3}
\end{figure}
%%%%%%%%%%%%%%%%%%%%%%%%%%%%%%%%%%%%%%%%%%%%%%%%%

%%%%%%%%%%%%%%%%%%%%%%%%%%%%%%%%%%%%%%%%%%%%%%%%%%%%%%%%%%%%%%%%%%%%%%%

\begin{figure}[h!]\centering
{\includegraphics[width=1.\textwidth]{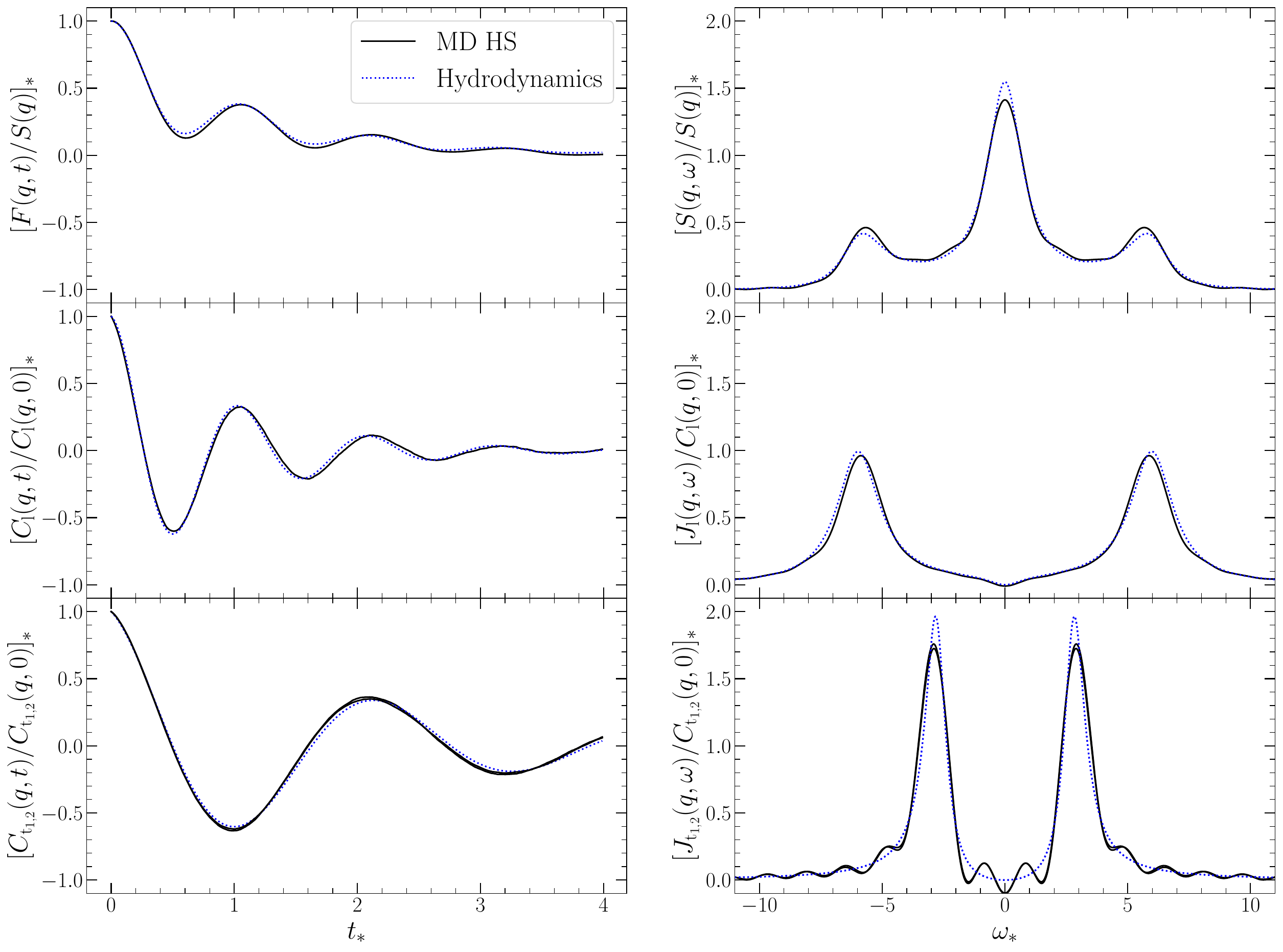}}
\caption[] {Normalized correlation and spectral functions at density $n_*=1.037$ and $q_*=0.50$ in the direction $[100]$ for a solid of $N=2048$ hard spheres. Left panel: Correlation functions versus time: From top to bottom: Intermediate scattering function $F({q},t)$, longitudinal component $C_{\rm l}({q},t)$, and transverse components $C_{{\rm t}_{1,2}}({q},t)$ of the time-dependent correlation functions, normalized by their values at $t=0$. Right panel: Spectral functions versus frequency:  From top to bottom: Dynamic structure factor $S({q},\omega)$, longitudinal component $J_{\rm l}({q},\omega)$, and transverse component $J_{{\rm t}_{1,2}}({q},\omega)$ of the spectral functions obtained from a numerical Fourier transform of the corresponding time-dependent correlation functions.  The dotted lines correspond to the analytical expressions predicted by hydrodynamics, given in equations~\eqref{eq:ISF} for $F({q},t)$,~\eqref{eq:lca_hydro} for $C_{\rm l}({q},t)$,~\eqref{eq:Ct} for $C_{{\rm t}_{1,2}}({q},t)$,~\eqref{eq:DSF_Full} for $S({q},\omega)$,~\eqref{eq:lsa_hydro} for $J_{\rm l}({q},\omega)$  and~\eqref{eq:JtoCt} for $J_{{\rm t}_{1,2}}({q},\omega)$.}\label{Fig:CF100-1.037}
\end{figure}

%%%%%%%%%%%%%%%%%%%%%%%%%%%%%%%%%%%%%%%%%%%%%%%%%%%%%%%%%%%%%%%%%%%%%%%

\begin{figure}[h!]\centering
{\includegraphics[width=1.\textwidth]{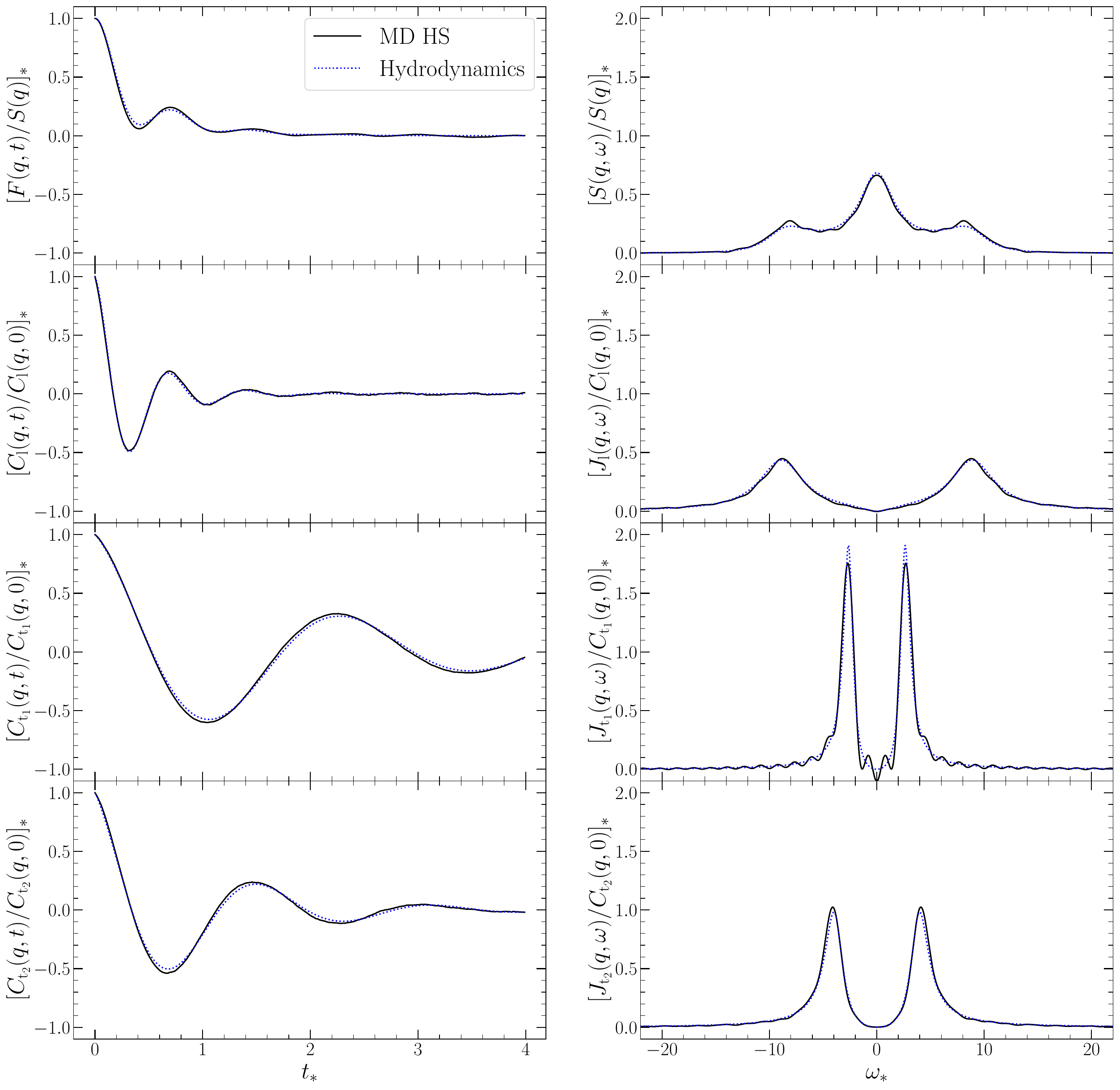}}
\caption[] {Normalized correlation and spectral functions at density $n_*=1.037$ and $q_*=0.71$ in the direction $[110]$ for a solid of $N=2048$ hard spheres. Left panel: Correlation functions versus time: From top to bottom: Intermediate scattering function $F({q},t)$, longitudinal component $C_{\rm l}({q},t)$, and transverse components $C_{{\rm t}_{1,2}}({q},t)$ of the time-dependent correlation functions, normalized by their values at $t=0$. Right panel: Spectral functions versus frequency:  From top to bottom: Dynamic structure factor $S({q},\omega)$, longitudinal component $J_{\rm l}({q},\omega)$, and transverse component $J_{{\rm t}_{1,2}}({q},\omega)$ of the spectral functions obtained from a numerical Fourier transform of the corresponding time-dependent correlation functions.  The dotted lines correspond to the analytical expressions predicted by hydrodynamics, given in equations~\eqref{eq:ISF} for $F({q},t)$,~\eqref{eq:lca_hydro} for $C_{\rm l}({q},t)$,~\eqref{eq:Ct} for $C_{{\rm t}_{1,2}}({q},t)$,~\eqref{eq:DSF_Full} for $S({q},\omega)$,~\eqref{eq:lsa_hydro} for $J_{\rm l}({q},\omega)$  and~\eqref{eq:JtoCt} for $J_{{\rm t}_{1,2}}({q},\omega)$.}\label{Fig:CF110-1.037}
\end{figure}

%%%%%%%%%%%%%%%%%%%%%%%%%%%%%%%%%%%%%%%%%%%%%%%%%%%%%%%%%%%%%%%%%%%%%%%
\begin{figure}[h!]\centering
{\includegraphics[width=1.\textwidth]{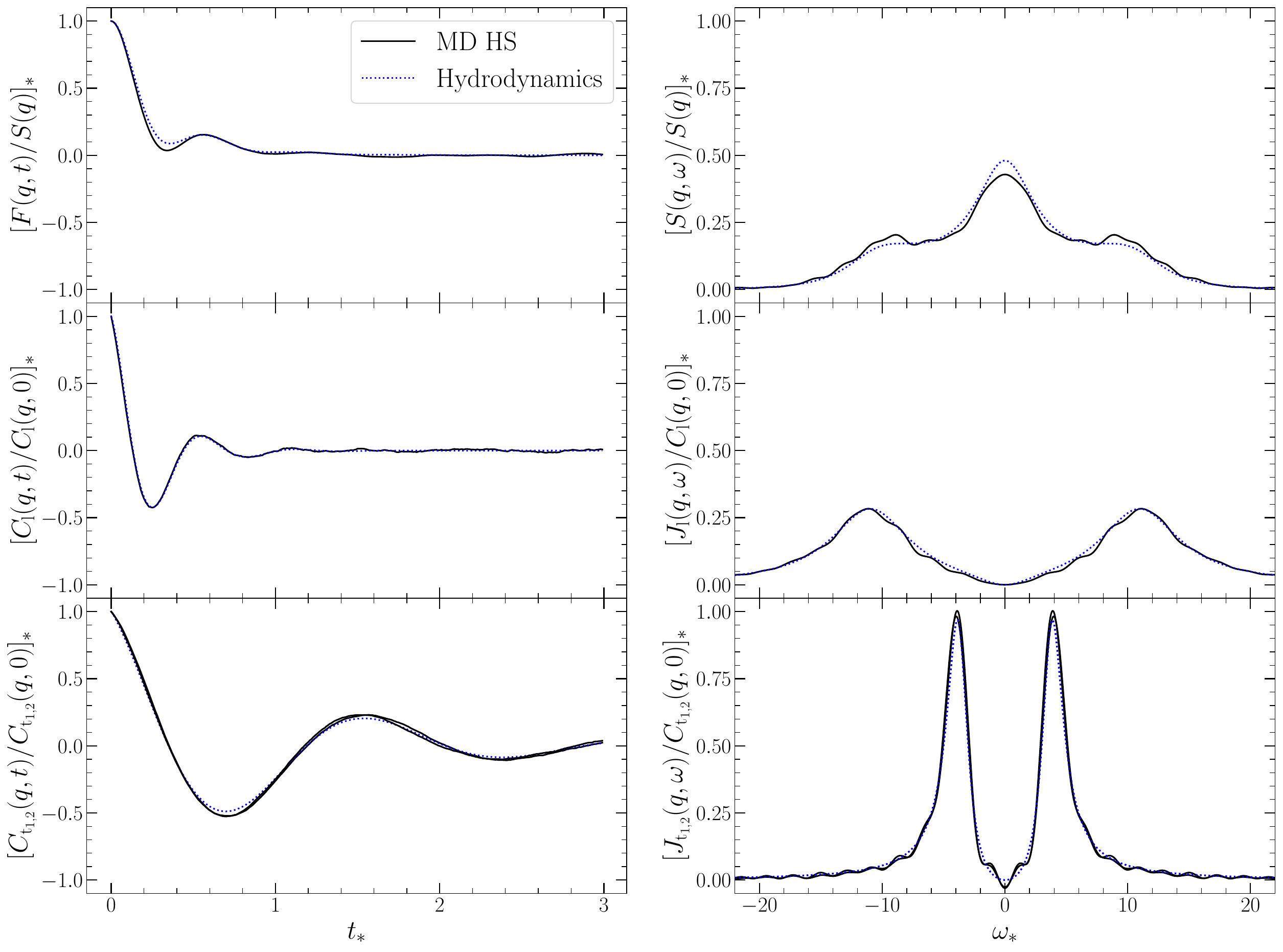}}
\caption[] {Normalized correlation and spectral functions at density $n_*=1.037$ and $q_*=0.87$ in the direction $[111]$ for a solid of $N=2048$ hard spheres. Left panel: Correlation functions versus time: From top to bottom: Intermediate scattering function $F({q},t)$, longitudinal component $C_{\rm l}({q},t)$, and transverse components $C_{{\rm t}_{1,2}}({q},t)$ of the time-dependent correlation functions, normalized by their values at $t=0$. Right panel: Spectral functions versus frequency:  From top to bottom: Dynamic structure factor $S({q},\omega)$, longitudinal component $J_{\rm l}({q},\omega)$, and transverse component $J_{{\rm t}_{1,2}}({q},\omega)$ of the spectral functions obtained from a numerical Fourier transform of the corresponding time-dependent correlation functions.  The dotted lines correspond to the analytical expressions predicted by hydrodynamics, given in equations~\eqref{eq:ISF} for $F({q},t)$,~\eqref{eq:lca_hydro} for $C_{\rm l}({q},t)$,~\eqref{eq:Ct} for $C_{{\rm t}_{1,2}}({q},t)$,~\eqref{eq:DSF_Full} for $S({q},\omega)$,~\eqref{eq:lsa_hydro} for $J_{\rm l}({q},\omega)$  and~\eqref{eq:JtoCt} for $J_{{\rm t}_{1,2}}({q},\omega)$.}\label{Fig:CF111-1.037}
\end{figure}

%%%%%%%%%%%%%%%%%%%%%%%%%%%%%%%%%%%%%%%%%%%%%%%%%%%%%%%%%%%%%%%%%%%%%%%
\begin{figure}[h!]\centering
{\includegraphics[width=1.\textwidth]{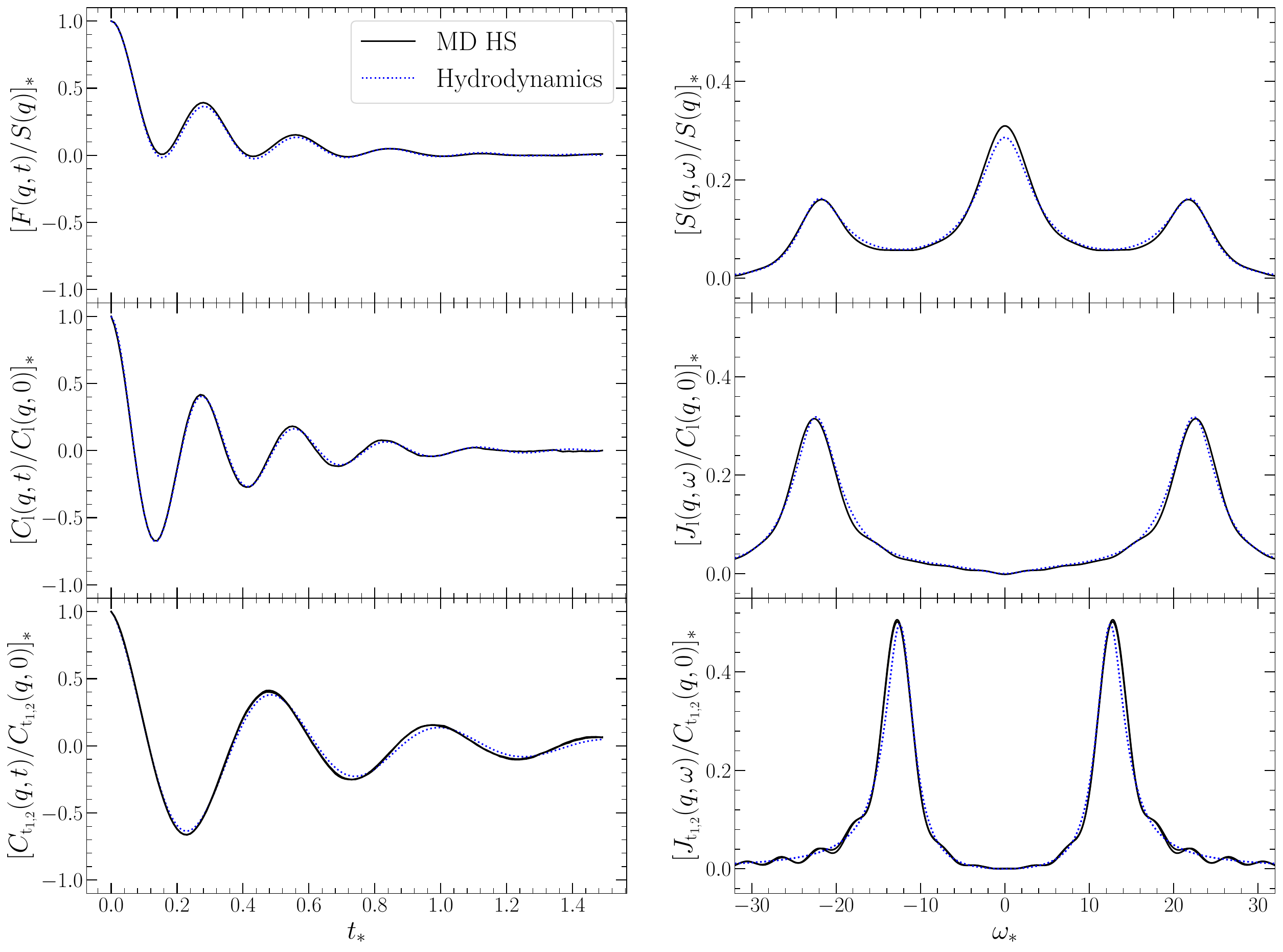}}
\caption[] {Normalized correlation and spectral functions at density $n_*=1.3$ and $q_*=0.54$ in the direction $[100]$ for a solid of $N=2048$ hard spheres. Left panel: Correlation functions versus time: From top to bottom: Intermediate scattering function $F({q},t)$, longitudinal component $C_{\rm l}({q},t)$, and transverse components $C_{{\rm t}_{1,2}}({q},t)$ of the time-dependent correlation functions, normalized by their values at $t=0$. Right panel: Spectral functions versus frequency:  From top to bottom: Dynamic structure factor $S({q},\omega)$, longitudinal component $J_{\rm l}({q},\omega)$, and transverse component $J_{{\rm t}_{1,2}}({q},\omega)$ of the spectral functions obtained from a numerical Fourier transform of the corresponding time-dependent correlation functions.  The dotted lines correspond to the analytical expressions predicted by hydrodynamics, given in equations~\eqref{eq:ISF} for $F({q},t)$,~\eqref{eq:lca_hydro} for $C_{\rm l}({q},t)$,~\eqref{eq:Ct} for $C_{{\rm t}_{1,2}}({q},t)$,~\eqref{eq:DSF_Full} for $S({q},\omega)$,~\eqref{eq:lsa_hydro} for $J_{\rm l}({q},\omega)$  and~\eqref{eq:JtoCt} for $J_{{\rm t}_{1,2}}({q},\omega)$.}\label{Fig:CF100-1.3}
\end{figure}

%%%%%%%%%%%%%%%%%%%%%%%%%%%%%%%%%%%%%%%%%%%%%%%%%%%%%%%%%%%%%%%%%%%%%%%
\begin{figure}[h!]\centering
{\includegraphics[width=1.\textwidth]{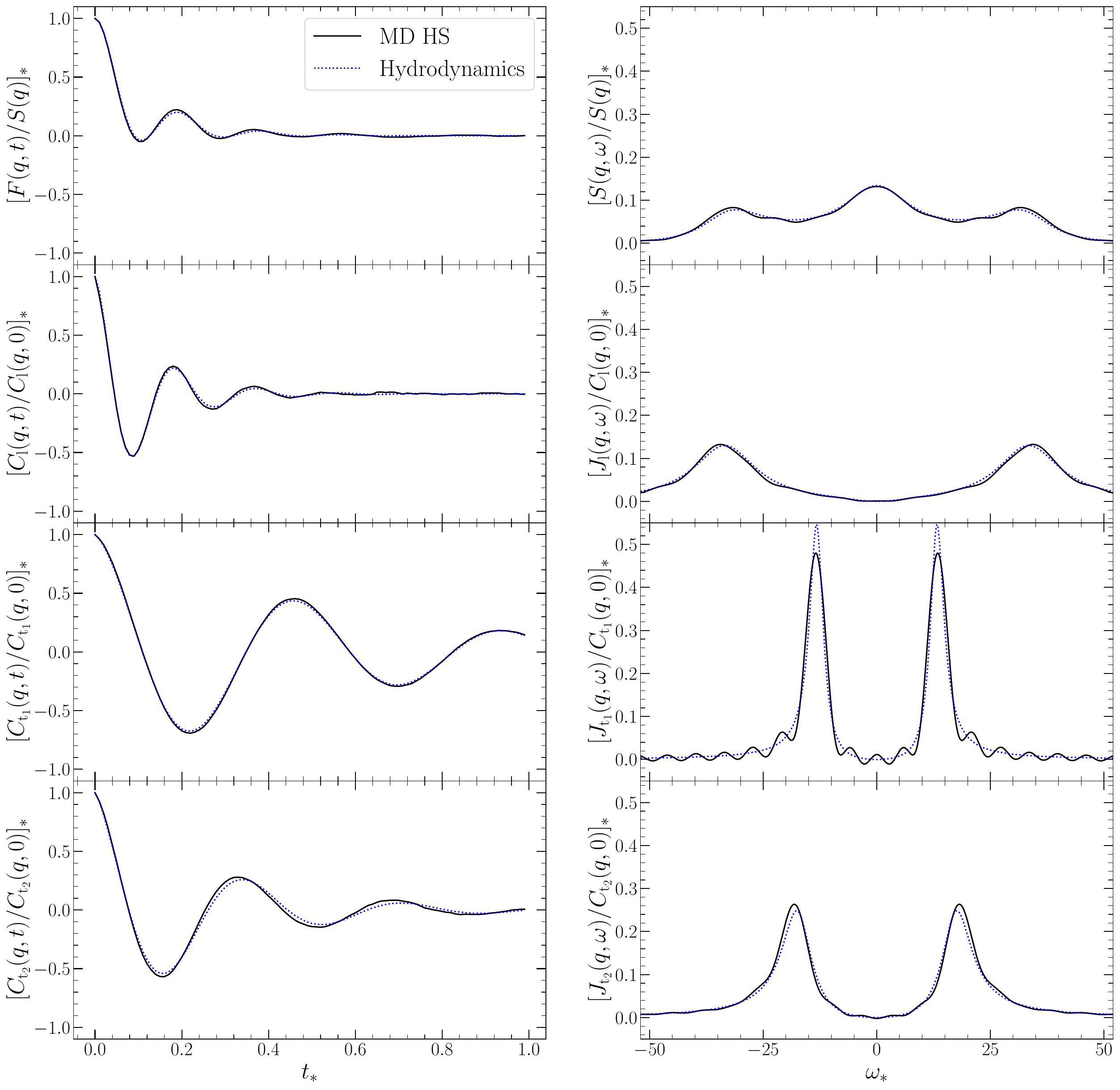}}
\caption[] {Normalized correlation and spectral functions at density $n_*=1.3$ and $q_*=0.76$ in the direction $[110]$ for a solid of $N=2048$ hard spheres. Left panel: Correlation functions versus time: From top to bottom: Intermediate scattering function $F({q},t)$, longitudinal component $C_{\rm l}({q},t)$, and transverse components $C_{{\rm t}_{1,2}}({q},t)$ of the time-dependent correlation functions, normalized by their values at $t=0$. Right panel: Spectral functions versus frequency:  From top to bottom: Dynamic structure factor $S({q},\omega)$, longitudinal component $J_{\rm l}({q},\omega)$, and transverse component $J_{{\rm t}_{1,2}}({q},\omega)$ of the spectral functions obtained from a numerical Fourier transform of the corresponding time-dependent correlation functions.  The dotted lines correspond to the analytical expressions predicted by hydrodynamics, given in equations~\eqref{eq:ISF} for $F({q},t)$,~\eqref{eq:lca_hydro} for $C_{\rm l}({q},t)$,~\eqref{eq:Ct} for $C_{{\rm t}_{1,2}}({q},t)$,~\eqref{eq:DSF_Full} for $S({q},\omega)$,~\eqref{eq:lsa_hydro} for $J_{\rm l}({q},\omega)$  and~\eqref{eq:JtoCt} for $J_{{\rm t}_{1,2}}({q},\omega)$.}\label{Fig:CF110-1.3}
\end{figure}

%%%%%%%%%%%%%%%%%%%%%%%%%%%%%%%%%%%%%%%%%%%%%%%%%%%%%%%%%%%%%%%%%%%%%%%

\begin{figure}[h!]\centering
{\includegraphics[width=1.\textwidth]{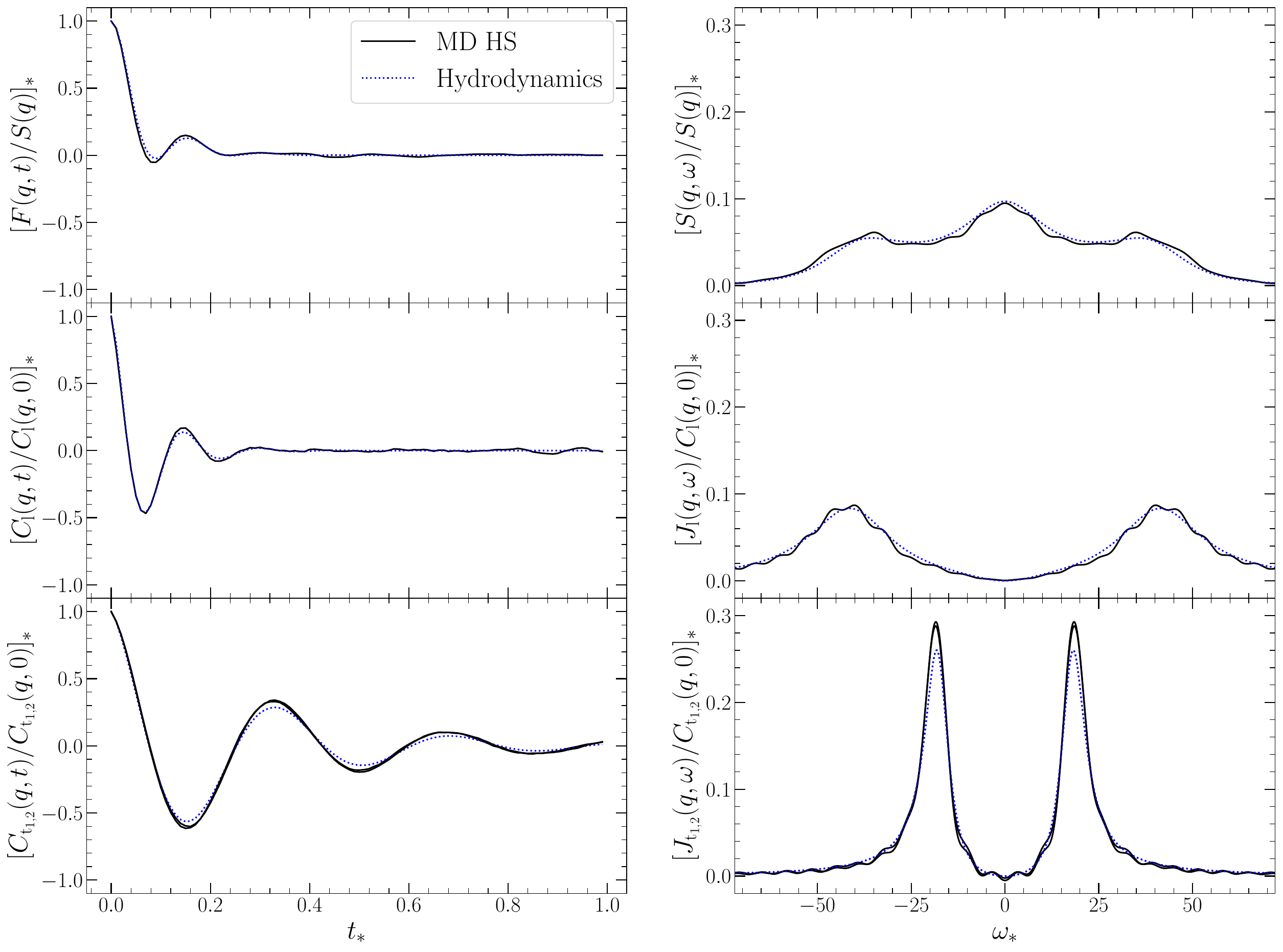}}
\caption[] {Normalized correlation and spectral functions at density $n_*=1.3$ and $q_*=0.93$ in the direction $[111]$ for a solid of $N=2048$ hard spheres. Left panel: Correlation functions versus time: From top to bottom: Intermediate scattering function $F({q},t)$, longitudinal component $C_{\rm l}({q},t)$, and transverse components $C_{{\rm t}_{1,2}}({q},t)$ of the time-dependent correlation functions, normalized by their values at $t=0$. Right panel: Spectral functions versus frequency:  From top to bottom: Dynamic structure factor $S({q},\omega)$, longitudinal component $J_{\rm l}({q},\omega)$, and transverse component $J_{{\rm t}_{1,2}}({q},\omega)$ of the spectral functions obtained from a numerical Fourier transform of the corresponding time-dependent correlation functions. The dotted lines correspond to the analytical expressions predicted by hydrodynamics, given in equations~\eqref{eq:ISF} for $F({q},t)$,~\eqref{eq:lca_hydro} for $C_{\rm l}({q},t)$,~\eqref{eq:Ct} for $C_{{\rm t}_{1,2}}({q},t)$,~\eqref{eq:DSF_Full} for $S({q},\omega)$,~\eqref{eq:lsa_hydro} for $J_{\rm l}({q},\omega)$  and~\eqref{eq:JtoCt} for $J_{{\rm t}_{1,2}}({q},\omega)$.}\label{Fig:CF111-1.3}
\end{figure}

%%%%%%%%%%%%%%%%%%%%%%%%%%%%%%%%%%%%%%%%%%%%%%%%%%%%%%%%%%%%%%%%%%%%%%%

\end{document}